\documentclass[preprint,review,12pt,fleqn]{elsarticle}

\usepackage{graphicx}
\usepackage{amssymb}
\usepackage{pgf}
\usepackage{tikz}
\usepackage{tabularx}
\usepackage{multirow}
\usepackage{subfigure}
\usepackage{placeins}
\usepackage{psfrag}
\usepackage{txfonts}

\journal{}

\definecolor{myblue}{rgb}{0.00,0.25,0.50}
\definecolor{mygreen}{rgb}{0.00,0.46,0.00}
\definecolor{myred}{rgb}{.750,0.0,0.00}

\begin{document}
\thicklines

\setlength{\baselineskip}{1.2\baselineskip}

\begin{frontmatter}

\title{Numerical simulation of the transient aerodynamic phenomena induced by passing manoeuvres}

\author[label1]{David Uystepruyst}
\author[label1]{Sini\v{s}a Krajnovi\'c}

\address[label1]{Division of Fluid Dynamics, Department of Applied Mechanics, Chalmers University of
Technology, SE-41296 Gothenburg, Sweden.}

\begin{abstract}
Several three-dimensional Unsteady Reynolds-Averaged Navier-Stokes (URANS) simulations of the passing generic vehicles (Ahmed bodies) are presented. The relative motion of vehicles was obtained using a combination of deforming and sliding computational grids. The vehicle studied is an Ahmed body with an angle of the rear end slanted surface of $30^{\circ}$. Several different relative velocities and transversal distances between vehicles were studied. The aerodynamic influence of the passage on the overtaken vehicle was studied. The results of the simulations were found to agree well with the existing experimental data. Numerical results were used to explain effects of the overtaking manoeuvre on the main aerodynamic coefficients.
\end{abstract}


\end{frontmatter}

\section{Introduction}
\label{intro}

The overtaking manoeuvre between two vehicles yields additional aerodynamic forces acting on both vehicles. These additional forces lead to sudden lateral displacements and rotations around the yaw axis of each vehicles. Such sudden change of the side force and of the yawing moment, complicates the steering corrections performed by the driver and can yield critical safety situations, in particular in adverse weather conditions, such as crosswinds or rain. Intensities of these forces are extrapolated when the overtaking manoeuvre involves a light car and a heavy-truck. Moreover, such a manoeuvre implies a disturbance of the aerodynamic of vehicles.\\

\noindent The first studies on the overtaking  effects, Heffrey \cite{Heffrey1973} and Howell \cite{Howell1973}, were investigated in response of the weight reduction of cars involved by the first oil crisis. Actually, after this oil crisis, car manufacturers have made substantial efforts to reduce the fuel consumption. This was achieved by improving the design of the cars, by developping efficient engine or by decreasing the vehicle weight. With the third option, vehicles became more sensitive to unsteady aerodynamic effects, such as those induced by an overtaking manoeuvre. In addition to the works of Heffrey and Howell, several experimental studies were carried out as well. These studies were focused on different aspects. Studies performed by Legouis et al \cite{Legouis1984}, Telionis et al \cite{Telionis1984} or Yamamoto et al \cite{Yamamoto1997} were dedicated to the car-truck overtaking. \\

\noindent More recently, several dynamic studies in order to analyze effects of the relative velocity, the transverse passing, and the crosswind during an overtaking manoeuvre were performed by Noger et al \cite{Noger2004,Noger2005}. Both studies were carried out with 7/10 scaled Ahmed bodies,  \cite{Ahmed1984}. The two bodies of the first study were hatchback shapes (slant of $30^{\circ}$), while the two bodies of the second one were squareback shapes. Noger and Van Grevenynghe \cite{Noger2011} proposed a study of car-truck overtaking on one test case: one relative velocity and one lateral spacing. Gilli\'eron and Noger \cite{Gillieron2004} analyzed the transient phenomena occurring during various phenomena such as the overtaking, the crossing or tunnel exits. \\

\noindent The overtaking process has also been studied using numerical modelling. Some recent two-dimensional (2D) numerical studies can be found. Clark and Filippone \cite{Clarke2007} performed the overtaking process of two sharped edges bodies. The work aimed to provide a thorough analysis of the overtaking process. Effects of the relative velocity and the transversal spacing were studied.
The authors focused on 2D overtaking as a preliminary means of investigating an appropriate simulation strategy for the complex three-dimensional (3D) flow.
Corin et al \cite{Corin2008} performed 2D numerical simulations of two rounded edges bodies. The dynamic effect of the passing manoeuvre was highlighted by comparisons with quasi-steady calculations. It was shown that crosswinds yield significant dynamic effects. The authors of \cite{Clarke2007,Corin2008} agreed that their 2D approaches were first steps towards 3D calculations. In particular, the Venturi effect, occurring when vehicles move closer, is strongly overestimated. 3D numerical simulations of two Ahmed bodies overtaking were performed by Gilli\'eron \cite{Gillieron2003}. Calculations were achieved using a Reynolds Averaged Navier-Stokes numerical method with a $k-\epsilon$ turbulence model. The effects of the transversal spacing and the crosswind were studied. However, this study was limited to a quasi-steady approach. \\

\noindent This paper presents a dynamic three-dimensional simulation of passing processes based on the $\zeta-f$ turbulence model and a deforming/sliding mesh method. The aims are to accurately predict the aerodynamic forces and moment, occurring on vehicles during a passing manoeuvre, and to give thorough analysis of the passing process. The paper starts with a description in section \ref{ssec:setup} of the experimental set-up that is used in the present numerical study. This is followed by the numerical methodology, including the turbulence model, the numerical method and details, and the deforming/sliding mesh method, in section \ref{ssec:numdet}. Results of the simulations are presented in section \ref{resultats}. In section \ref{sec:discus}, a complete analysis of the effects of the passing manoeuvre is provided. Finally, the paper is summarized in section \ref{sec:conc}.

\section{Method}
\label{sec:met}

\subsection{Description of the set-up}
\label{ssec:setup}

\subsubsection{Geometries}
\label{sssec:geom}

\begin{figure}
\centering
\begin{tikzpicture}[scale=0.1]
\draw[dashed] (0,-2) -- (0,32); \draw[dashed] (73.08,-2) -- (73.08,32);
\draw[dashed] (14.14,-2) -- (14.14,5.5); \draw[dashed] (47.04,-2) -- (47.04,5.5);
\draw[->] (36.54,32) -- (0,32); \draw[->] (36.54,32) -- (73.08,32); \node at (36.54,33.5) {\tiny 730.8};
\draw[dashed] (73.08,5.5) -- (77,5.5); \draw[dashed] (73.08,17.32) -- (77,17.32);
\draw[->] (77,11.16) -- (77,17.32); \draw[->] (77,11.16) -- (77,5.5); \node at (75.5,11.16) {\tiny \rotatebox{90}{123.2}};
\draw[->] (7.09,-2) -- (0,-2); \draw[->] (7.09,-2) -- (14.14,-2); \node at (7.09,-1) {\tiny 141.4};
\draw[->] (30.59,-2) -- (14.14,-2); \draw[->] (30.59,-2) -- (47.04,-2); \node at (30.59,-1) {\tiny 329};
\draw[->] (60.06,-2) -- (73.08,-2); \draw[->] (60.06,-2) -- (47.04,-2); \node at (60.06,-1) {\tiny 260.4};
\draw[thick] (7,5.5) -- (73.08,5.5) -- (73.08,17.32) -- (59.5,25.66) -- (7,25.66); \draw[thick] (7,18.66) +(90:7) arc (90:180:7);  \draw[thick] (0,5.5) +(90:7) arc (180:270:7); \draw[thick] (0,12.5) -- (0,18.66); \draw[dashed] (59.5,25.66) -- (70,25.66);
\draw[thick] (13.39,5.5) -- (13.39,0) -- (14.89,0) -- (14.89,5.5);  \draw[thick] (46.29,5.5) -- (46.29,0) -- (47.79,0) -- (47.79,5.5);
\draw (7,5.5) -- (7,25.66);
\draw[dashed] (90,6) -- (90,32); \draw[dashed] (117.23,6) -- (117.23,32);
\draw[->] (103.615,32) -- (90,32); \draw[->] (103.615,32) -- (117.23,32); \node at (103.615,33) {\tiny 272.3};
\draw[dashed] (92.165,5.5) -- (92.165,-2); \draw[dashed] (115.065,5.5) -- (115.065,-2);
\draw[->] (103.615,-2) -- (92.165,-2); \draw[->] (103.615,-2) -- (115.065,-2); \node at (103.615,-1) {\tiny 229};
\draw[dashed] (90,0) -- (86,0);
\draw[->] (86,2.75) -- (86,5.5); \draw[->] (86,2.75) -- (86,0); \node at (87.5,2.75) {\tiny \rotatebox{90}{55}};
\draw[dashed] (90,5.5) -- (86,5.5); \draw[dashed] (90,25.66) -- (86,25.66);
\draw[->] (86,15.58) -- (86,5.5); \draw[->] (86,15.58) -- (86,25.66); \node at (87.5,15.58) {\tiny \rotatebox{90}{201.6}};
\draw[thick] (90,5.5) -- (90,25.66) -- (117.23,25.66) -- (117.23,5.5) -- (90,5.5); \draw[thick] (91.415,5.5) -- (91.415,0) -- (92.915,0) -- (92.915,5.5);
\draw[thick] (114.315,5.5) -- (114.315,0) -- (115.815,0) -- (115.815,5.5); \draw[thick] (90,17.32) -- (117.23,17.32);
\draw[] (7,18.66) -- (2,23.66); \node at (3.5,19.5) {\tiny R70};
\draw (61.5,21.16) +(50:3) arc (310:360:3); \node at (68,24) {\tiny $30^{\circ}$};
\end{tikzpicture}
\caption{7/10 scale Ahmed bluff body \cite{Ahmed1984}, dimensions in $mm$.}
\label{ahmed_body}
\end{figure}
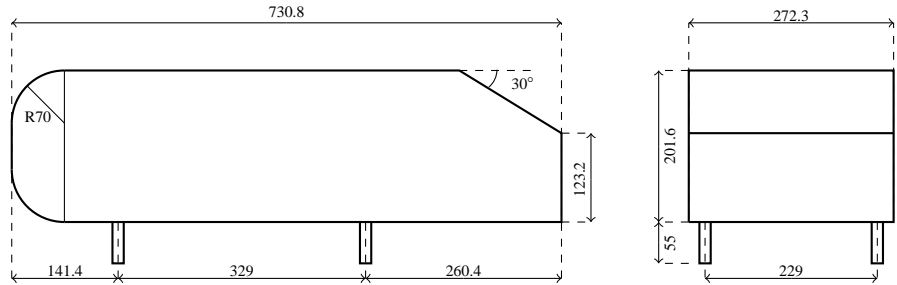

The body used in this study is identical to this in the experimental work \cite{Noger2005} and is 7/10 Ahmed bodies shown in figure \ref{ahmed_body}. This body has a hatchback type rear end with an angle of $30^{\circ}$. \\

\noindent As it is shown, both bodies consist on rounded front end and sharped rear end. The main vehicle sizes are: the length $L=730.8~mm$, the width $W=272.3~mm$, the height $H=201.6~mm$ and the ground clearance of $55~mm$. Supports are $15~mm$ diameter cylindrical and the length of the slant surface is $155.4~mm$. The Reynolds number based on height of the vehicle is $Re_{H}=390.000$, for a velocity of $30~m.s^{-1}$.

\subsubsection{Dimensionless coefficient}
\label{sssec:adim}

As in the experimental works of Noger et al, a dimensionless parameter $k$ is defined as the ratio of the relative velocity $V_{r}$ to a steady velocity $V$:
\begin{equation}
k=\displaystyle\frac{V_{r}}{V}.
\label{kv}
\end{equation}
The steady velocity is the velocity of the moving body, e.g. $V=V_{\infty}+V_{r}$. \\

\noindent During an overtaking, the strongly affected aerodynamic coefficients are the drag force coefficient $C_{d}$, the side force coefficient $C_{y}$ and the yawing moment coefficient $C_{n}$ given by:
\begin{equation}
\displaystyle
\left\{
\begin{array}{l}
 F_{d}=\displaystyle\frac{1}{2}\rho SV^{2}C_{d},\vspace{0.1cm}\\
 F_{y}=\displaystyle\frac{1}{2}\rho SV^{2}C_{y},\vspace{0.1cm}\\
 N=\displaystyle\frac{1}{2}\rho ESV^{2}C_{n},\\
\end{array}
\right.
\label{coeffs}
\end{equation}
where $F_{d}$, $F_{y}$ and $N$ are respectively the drag force, the side force and the yawing moment obtained by integrating the pressure distribution around the model. In equation (\ref{coeffs}), $\rho$ is the air density, $S$ the body frontal area and $E$ the longitudinal distance between the supports. Note that $V$ is the above defined steady velocity.

\subsubsection{Overtaking process}
\label{sssec:overproc}

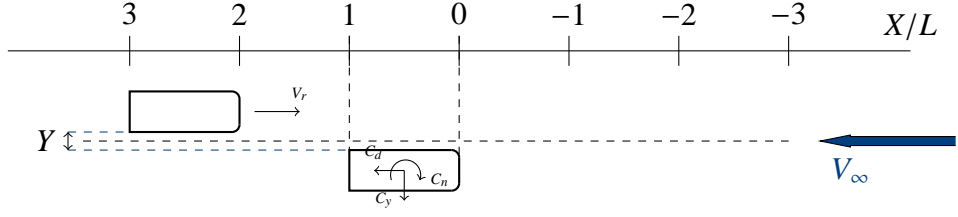
\begin{figure}
\centering
\begin{tikzpicture}[scale=2]
\draw[thick] (-0.05,-0.33) -- (-0.73,-0.33) -- (-0.73,-0.06) -- (-0.05,-0.06); \draw[thick] (-0.05,-0.33) arc (270:360:0.05); \draw[thick] (0,-0.11) arc (0:90:0.05); \draw[thick] (0,-0.11) -- (0,-0.28); \draw[->]  (-0.365,-0.19615) -- (-0.365,-0.39615) node[left] {\tiny $C_{y}$}; \draw[->]  (-0.365,-0.19615) -- (-0.565,-0.19615) node[above] {\tiny $C_{d}$};
\draw[->] (-0.45,-0.26) arc (200:-20:0.1) node[right] {\tiny $C_{n}$};
\draw[thick] (-1.51,0.33) -- (-2.19,0.33) -- (-2.19,0.06) -- (-1.51,0.06); \draw[thick] (-1.46,0.28) arc (0:90:0.05); \draw[thick] (-1.51,0.06) arc (270:360:0.05); \draw[thick] (-1.46,0.11) -- (-1.46,0.28);
\draw[->]  (-1.36,0.19615) -- (-1.06,0.19615) node[above] {\tiny $V_{r}$};
\filldraw[fill=myblue,draw=black]  (3.3,-0.03) -- (2.6,-0.03) --  (2.6,-0.04) -- (2.4,0) -- (2.6,0.04) -- (2.6,0.03) -- (3.3,0.03); \node[myblue] at (2.6,-0.2) {$V_{\infty}$};
\draw  (-3,0.6) -- (3,0.6) node[above] {$X/L$}; \draw[dashed]  (-2.5,0) -- (2.19,0);
\draw[dashed,thin,myblue]  (-2.6,0.06) -- (-2.19,0.06); \draw[dashed,thin,myblue]  (-2.6,-0.06) -- (-0.7308,-0.06);
\draw[->]  (-2.6,0) -- (-2.6,0.06); \draw[->]  (-2.6,0) -- (-2.6,-0.06); \node at (-2.75,0) {$Y$};
\draw (-2.19,0.5) -- (-2.19,0.7) node[above] {$3$}; 
\draw (-1.46,0.5) -- (-1.46,0.7) node[above] {$2$}; \draw (-0.73,0.5) -- (-0.73,0.7) node[above] {$1$};
\draw (0,0.5) -- (0,0.7) node[above] {$0$};
\draw (2.19,0.5) -- (2.19,0.7) node[above] {$-3$}; 
\draw (1.46,0.5) -- (1.46,0.7) node[above] {$-2$}; \draw (0.73,0.5) -- (0.73,0.7) node[above] {$-1$};
\draw[dashed] (-0.73,0.7) -- (-0.73,-0.1); \draw[dashed] (0,0.7) -- (0,-0.1);
\end{tikzpicture}
\caption{Notations for the vehicle positioning and aerodynamic coefficient direction.}
\label{over}
\end{figure}

The figure \ref{over} shows the sketch of the overtaking, with distances and the forces direction. The overtaking consists, as in the experimental work, on the stationnary body located in the middle of the wind tunnel length and a moving body located 5L behind the stationary body at the beginning of the calculation, and 5L in front of the stationary body at the end of the calculation. An inlet condition is set with the normal velocity $V_{\infty}$ corresponding to the velocity of the stationnary body. The moving body is set in motion with the relative velocity $V_{r}$. The reference case is set with the velocity ratio $k$ and the transversal spacing $Y$ at values $k=0.248$ ($V_{\infty}=30.32~m.s^{-1}$ and $V_{r}=10~m.s^{-1}$) and $Y=0.25W$.

\subsection{Numerical methodology}
\label{ssec:numdet}

\subsubsection{Governing equations and turbulence model}
\label{equa}

The flow around vehicles was predicted using the Reynolds-Averaged Navier Stokes equations coupled with the eddy viscosity $\zeta-f$ model equations \cite{Hanjalic2004}. The continuity and momentum equations are given by:

\begin{equation}
\left\{
\begin{array}{l}
 \displaystyle\frac{\partial U_{i}}{\partial x_{i}}=0,\vspace{.1cm}\\
 \displaystyle\frac{\partial U_{i}}{\partial t}+U_{j}\displaystyle\frac{\partial U_{i}}{\partial x_{j}}=-\displaystyle\frac{1}{\rho}\frac{\partial P}{\partial x_{i}}+\displaystyle\frac{1}{\rho}\frac{\partial}{\partial x_{j}}\left(\tau_{ij}-\rho\displaystyle\overline{u_{i}u_{j}}\right),\\
\end{array}
\right.
\label{ns2}
\end{equation}
where $U_{i}$ is the mean-velocity vector, $\rho$ is the fluid density, $P$ the mean-pressure, $\tau_{ij}$ denotes the mean viscous stress tensor:
\begin{equation}
\tau_{ij}=2\mu S_{ij}.
\label{tau}
\end{equation}
In the equation (\ref{tau}), $\mu$ is the dynamic viscosity and the mean strain rate tensor $S_{ij}$ is given by:
\[
\begin{array}{l}
S_{ij}=\displaystyle\frac{1}{2}\left(\displaystyle\frac{\partial U_{i}}{\partial x_{j}}+\displaystyle\frac{\partial U_{j}}{\partial x_{i}}\right).
\end{array}
\]

\noindent The last term of equation (\ref{ns2}) is the unknown Reynolds stress tensor which must be modeled. 

\noindent The Reynolds stress tensor is expressed with the Boussinesq's analogy:
\[
-\rho\displaystyle\overline{u_{i}u_{j}}=2\rho\nu_{t}S_{ij}-\displaystyle\frac{2}{3}\rho k\delta_{ij},
\]
where $\nu_{t}$ is the turbulent viscosity and $\delta_{ij}$ is the Kronecker delta. In the $\zeta-f$ model, the eddy-viscosity is defined as:
\[
\nu_{t}=C_{\mu}\zeta k \tau,
\]
where $\tau$ is the time scale given as:
\[
\tau=\max\left[\min\left(\displaystyle\frac{k}{\varepsilon},\displaystyle\frac{a}{\sqrt{6}C_{\mu}\left|S\right|\zeta}\right),C_{\tau}\left(\displaystyle\frac{\nu}{\varepsilon}\right)^{1/2}\right].
\]
The velocity scale ratio $\zeta$ is obtained from the following equation:
\[
\displaystyle\frac{D\zeta}{Dt}=f-\displaystyle\frac{\zeta}{k}\mathcal{P}_{k}+\displaystyle\frac{\partial}{\partial x_{k}}\left[\left(\nu+\displaystyle\frac{\nu_{t}}{\sigma_{\zeta}}\right)\displaystyle\frac{\partial\zeta}{\partial x_{k}}\right].
\]
The equations of the turbulent kinetic energy and its dissipation are:
\[
\displaystyle\frac{Dk}{Dt}=\left(\mathcal{P}_{k}-\varepsilon\right)+\displaystyle\frac{\partial}{\partial x_{j}}\left[\left(\nu+\displaystyle\frac{\nu_{t}}{\sigma_{k}}\right)\displaystyle\frac{\partial k}{\partial x_{j}}\right]
\]
\[
\displaystyle\frac{D\varepsilon}{Dt}=\displaystyle\frac{C_{\varepsilon1}\mathcal{P}_{k}-C_{\varepsilon2}\varepsilon}{\tau}+\displaystyle\frac{\partial}{\partial x_{j}}\left[\left(\nu+\displaystyle\frac{\nu_{t}}{\sigma_{\varepsilon}}\right)\displaystyle\frac{\partial\varepsilon}{\partial x_{j}}\right].
\]

\noindent In above equations, the production is given by:
\[
\mathcal{P}_{k}=-\overline{u_{i}u_{j}}\displaystyle\frac{\partial U_{i}}{\partial x_{j}}.
\]
The elliptic relaxation function $f$ is formulated by using the pressure-strain model of Speziale et al \cite{Speziale1991}: 
\[
L^{2}\nabla^{2}f-f=\displaystyle\frac{1}{\tau}\left(c_{1}+C_{2}^{\prime}\displaystyle\frac{\mathcal{P}_{k}}{\epsilon}\right)\left(\zeta-\displaystyle\frac{2}{3}\right)-\left(\displaystyle\frac{C_{4}}{3}-C_{5}\right)\displaystyle\frac{\mathcal{P}_{k}}{k}.
\]

\noindent The length scale $L$ is:
\[
L=C_{L}\max\left[\min\left(\displaystyle\frac{k^{3/2}}{\varepsilon},\displaystyle\frac{k^{1/2}}{\sqrt{6}C_{\mu}\left|S\right|\zeta}\right),C_{\eta}\left(\displaystyle\frac{\nu^{3}}{\varepsilon}\right)^{1/4}\right].
\]

\noindent Coefficients in above equations are given in table \ref{coeffs_zeta}.

\begin{table}
\begin{tabular}{lllllllllll}
\hline
$C_{\mu}$ & $C_{\varepsilon1}$ & $C_{\varepsilon2}$ & $c_{1}$ & $C_{2}^{\prime}$ & $\sigma_{k}$ & $\sigma_{\varepsilon}$ & $\sigma_{\zeta}$ & $C_{\tau}$ & $C_{L}$ & $C_{\eta}$ \\
\hline
$0.22$ & $1.4(1+0.012/\zeta)$ & $1.9$ & $0.4$ & $0.65$ & $1$ & $1.3$ & $1.2$ & $6.0$ & $0.36$ & $85$ \\
\hline
\end{tabular}
\caption{Coefficients in the $\zeta-f$ turbulence model}\label{coeffs_zeta}
\end{table}

\noindent The $\zeta-f$ model by Hanjali\'c et al \cite{Hanjalic2004}, based on the $\overline{v^{2}}-f$ model of Durbin \cite{Durbin1991}, is very robust and more accurate than the simple two-equation eddy viscosity models.

\subsubsection{Numerical method}

The system of equations (\ref{ns2}) was solved using a commercial solver, AVL FIRE. This software is based on a cell-centered finite volume method. The momentum equations were discretized using a second-order upwind scheme. An implicit second-order scheme was used for the temporal discretization. The SIMPLE algorithm was used to couple the velocity and pressure fields. A collocated grid arrangement was employed.\\

\begin{figure}[!ht]
\centering
\includegraphics[width=0.3\textwidth,angle=90,bb=160 0 475 770,clip=true]{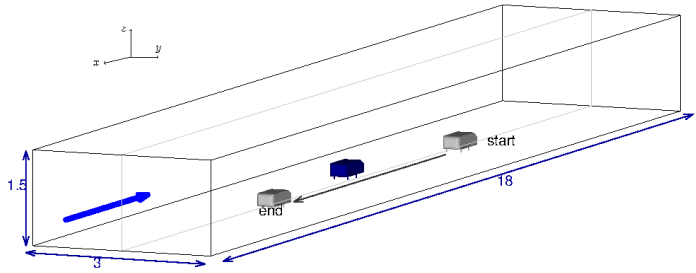}
\caption{Computational domain (dimensions in m).}
\label{domain}
\end{figure}

\noindent The numerical domain is shown in figure \ref{domain}. The experimental wind tunnel section is $5~m\times3~m$. This section was reduced in the present simulations. The width, and the height, of the numerical domain represent respectively more than 15 times, and 7.5 times, the height of the vehicle. The length is set to $18~m$ for the good progress of the deforming/sliding mesh strategy.\\

\noindent The uniform free stream velocity $V_{\infty}$ was set at the inlet boundary, in front of vehicles. A static pressure was applied at the outlet. No-slip wall boundary conditions were used on the bodies and on the floor. Finally, slip wall boundary conditions were applied on the lateral and on the roof surfaces.

\subsubsection{Numerical details}
\label{numdet2}

\begin{figure}[!ht]
\centering
\includegraphics[bb=105 0 450 750,clip=true,width=0.5\textwidth,angle=90]{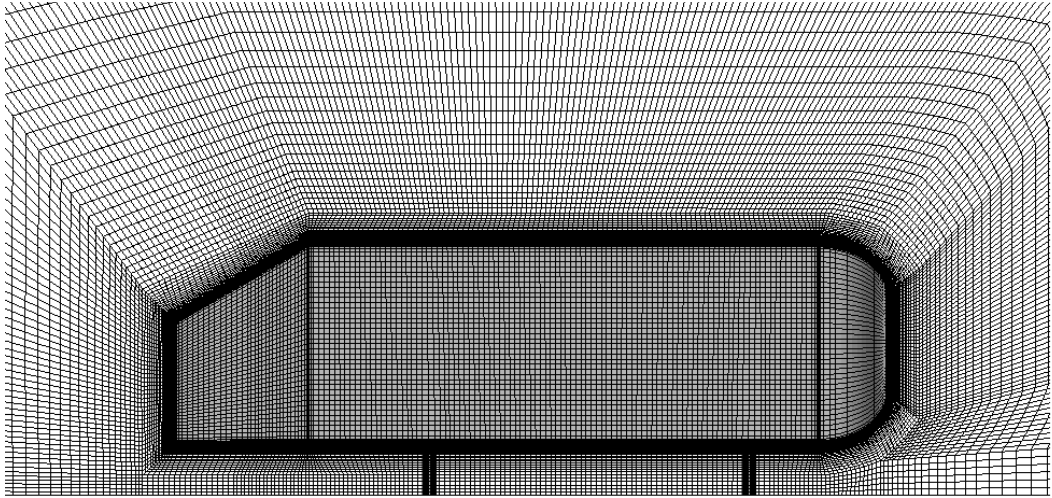}
\caption{Surface and volume mesh of bluff body.}
\label{side_mesh}
\end{figure}

The structured grids were made with the commercial grid generator Ansys ICEM-CFD and consist of only hexahedral elements. The figure \ref{side_mesh} shows a side view of volume and surface meshes, for the bluff body. A grid topology was constructed using several O-grids in order to concentrate most of the computational cells close to the surface of the vehicles.\\

\noindent Accuracy was established by making the original case simulation ($k=0.248$ and $Y=0.25W$) on three different computational grids. The numbers of computational cells are: 4 millions for the coarse mesh, 6 millions for the middle mesh and 8 millions for the fine mesh. For a velocity of $30~m.s^{-1}$, the wall normal resolution, $n^{+}$, is such that $n^{+}<2$, $n^{+}<4$ and $n^{+}<6$ for the fine mesh, the middle mesh and the coarse mesh, respectively. Note that the cell next to the wall should reach $n^{+}$ as a maximum less than 3 with the $\zeta-f$ model \cite{Krajnovic2009}. The resolution in the streamwise direction $\Delta s^{+}$, and the resolution in directions normal to streamwise $\Delta l^{+}$ are reported in table \ref{nodes_distrib} for the three computational meshes. The averaged $\Delta s^{+}$ was 450, 350 and 300 for the coarse, the middle and the fine grid, respectively. The averaged $\Delta l^{+}$ was 400, 300 and 260 for the coarse, the middle and the fine grid, respectively. Here, $n^{+}=nu_{\tau}/\nu$, $\Delta s^{+}=\Delta su_{\tau}/\nu$ and $\Delta l^{+}=\Delta lu_{\tau}/\nu$, where $n$ is the wall-normal distance, $\Delta s$ is the streamwise distance, $\Delta l$ is the spanwise distance, $u_{\tau}$ is the friction velocity and $\nu=\mu/\rho$ is the kinematic viscosity. The time step range was between $2\times10^{-4}~s$ and $5\times10^{-4}~s$, depending on the grid and the relative velocity, giving a CFL number around 0.9 for the highest velocity and for all grids.

\begin{table}
\begin{tabularx}{\textwidth}{XXX}
\hline
 coarse & middle & fine \\
\hline\rule[0ex]{0pt}{3ex}
 $20<\Delta s^{+}<960$ & $20<\Delta s^{+}<580$ & $20<\Delta s^{+}<500$ \\
\hline\rule[0ex]{0pt}{3ex}
 $20<\Delta l^{+}<840$ & $20<\Delta l^{+}<560$ & $20<\Delta l^{+}<450$ \\
\hline
\end{tabularx}
\caption{Resolution in the streamwise and normal streamwise directions.}\label{nodes_distrib}
\end{table}

\subsubsection{Deforming and sliding mesh}
\label{slidemesh}

\begin{figure}[!ht]
\centering
\subfigure[start]{\includegraphics[bb=90 70 600 200,clip=true,width=0.9\textwidth]{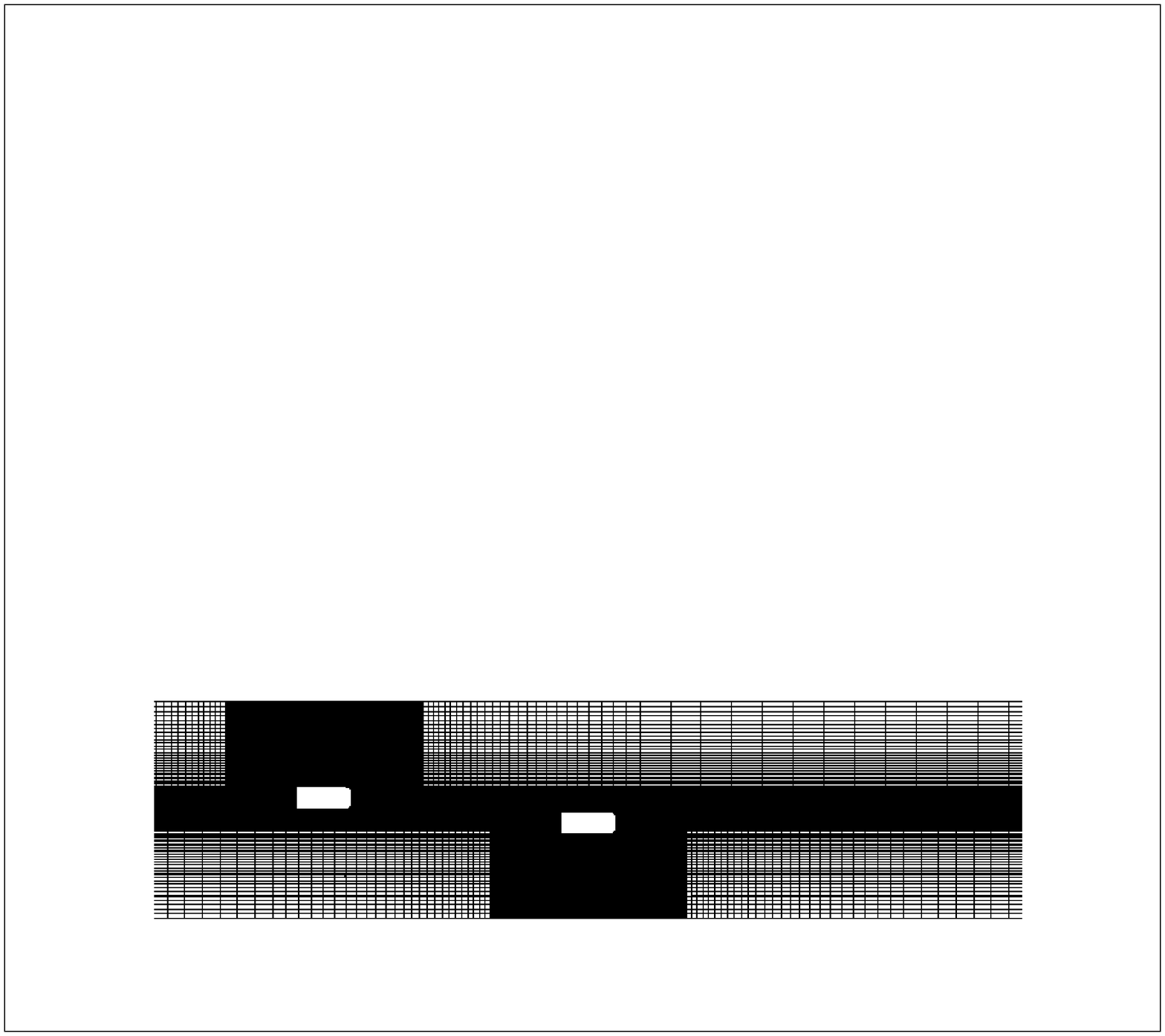}\label{deform_mesh1}}
\subfigure{\includegraphics[bb=90 70 600 205,clip=true,width=0.9\textwidth]{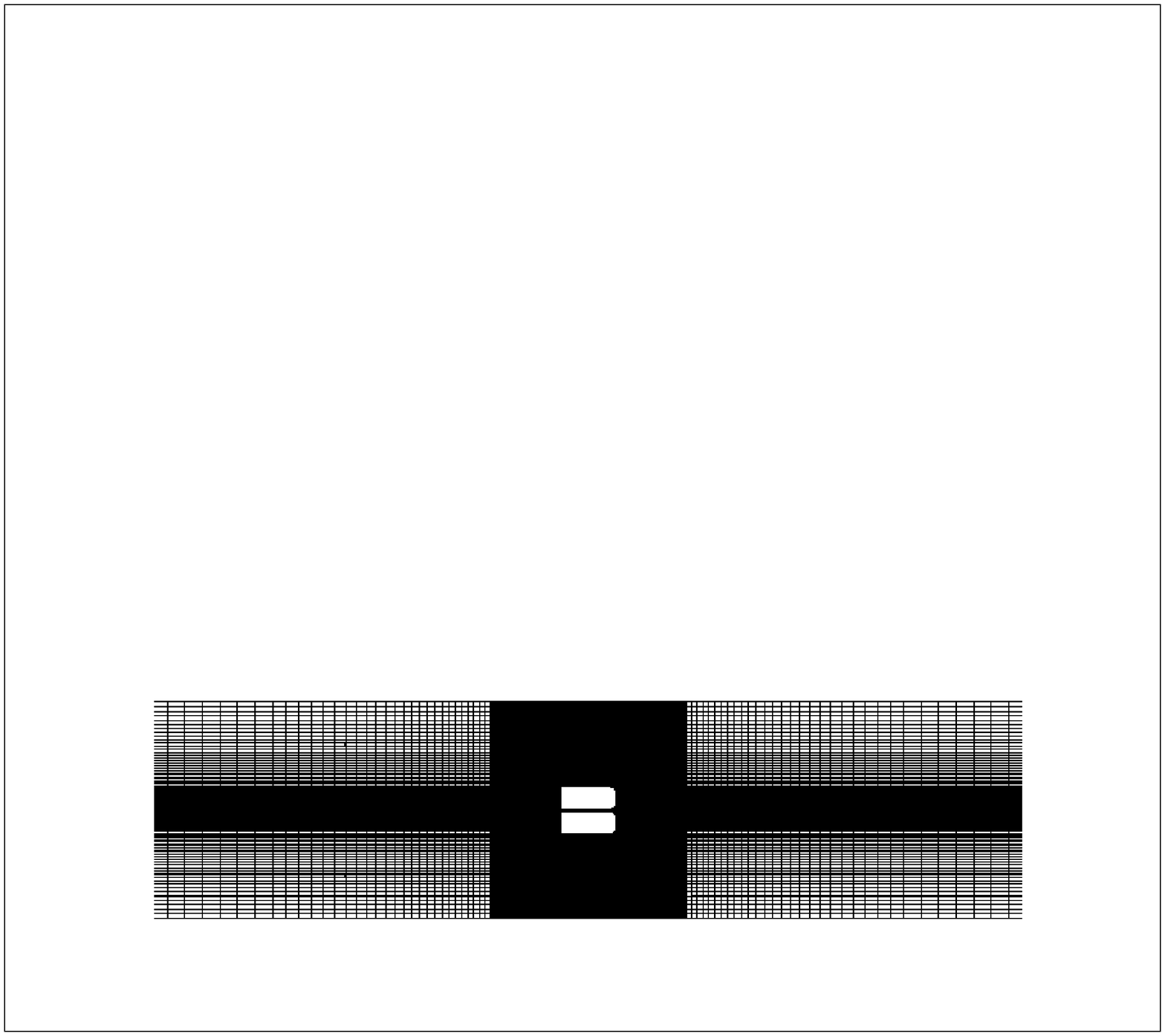}}
\setcounter{subfigure}{1}
\subfigure[middle]{
\centering
\begin{tikzpicture}[scale=0.1]
\draw[thick] (-0.7,.55) -- (-7.308,.55) -- (-7.308,2.566) -- (-.7,2.566); \draw[thick] (0,1.166) +(90:.7) arc (0:90:.7);  \draw[thick] (-.7,-0.15) +(90:.7) arc (270:360:.7); \draw[thick] (0,1.25) -- (0,1.866);
\draw[thick] (-1.339,.55) -- (-1.339,0) -- (-1.489,0) -- (-1.489,.55);  \draw[thick] (-4.629,.55) -- (-4.629,0) -- (-4.779,0) -- (-4.779,.55);
\draw (-63.808,0) -- (55.5,0); \draw[very thin, white] (61,0.00001) -- (61,-0.00001);
\draw (-63.808,1) -- (-63.808,-1) node [below] {$r_{1}$}; \draw (-22.308,1) -- (-22.308,-1) node [below] {$r_{2}$}; \draw (15.308,1) -- (15.308,-1) node [below] {$r_{3}$}; \draw (55.5,1) -- (55.5,-1) node [below] {$r_{4}$};
\end{tikzpicture}\label{deform_explan}}
\subfigure[end]{\includegraphics[bb=90 70 600 205,clip=true,width=0.9\textwidth]{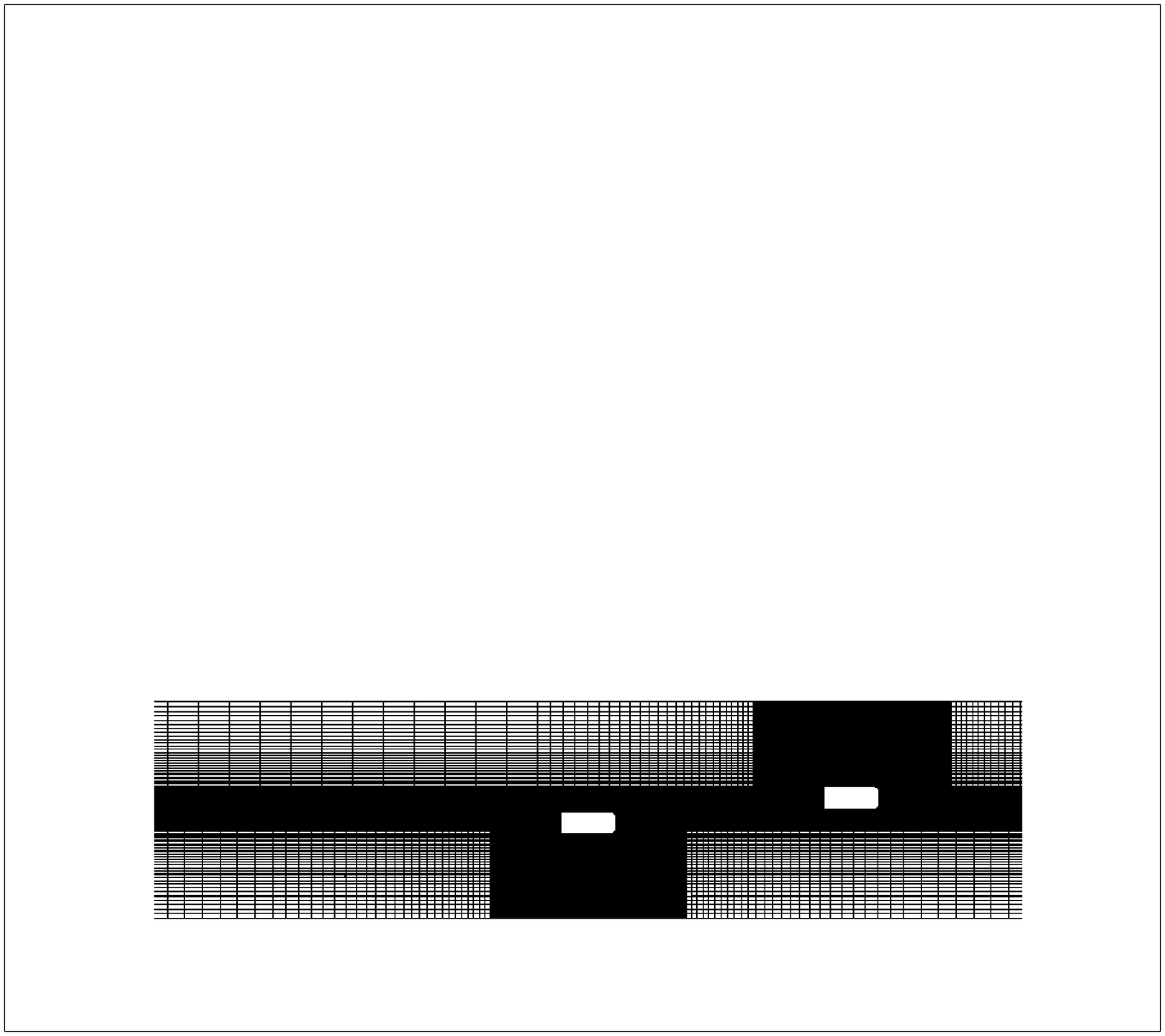}}
\caption{Deformation of the computational grid for the overtaking process.}
\label{deform_mesh}
\end{figure}

\noindent The rectilinear displacement of a body was achieved by a deforming/sliding grid method. Figure \ref{deform_mesh} illustrates the effect of the body movement on the computational grid at three different times of the simulation. \\

\noindent The overall domain was composed by two subdomains: the bottom one, containing the stationnary body, remained fixed all along the simulation; and the top one, containing the moving body. This last subdomain is symbolically divided in three zones $\left[r_{i},r_{i+1}\right]$, for $i=1,~2,~3$, see figure \ref{deform_explan}. The zone between $r_{2}$ and $r_{3}$ was slided during the simulation. The two remaining zones were compressed or stretched in response of the sliding movement of the central part. A similar approach was successfully used by Krajnovi\'c et al. \cite{Krajnovic2009,Krajnovic2011} for the unsteady RANS simulations of trains passing each other or exiting tunnel and for the LES simulations of a rotating vehicle.
Finally a common interfacing was performed between the two meshes.

\section{Results}
\label{resultats}

\subsection{Numerical accuracy}

\begin{figure}[!ht] 
\centering 
\subfigure{
 \includegraphics[width=\textwidth,bb=50 305 725 575,clip=true]{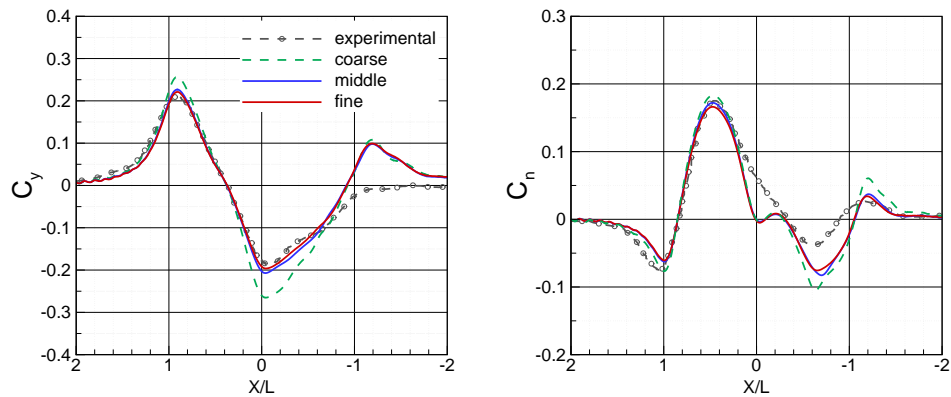}}
\caption{Mesh resolution analysis. Side force (left) and yawing moment (right) coefficients.}\label{mesh_resol}
\end{figure}

\noindent Numerical accuracy was checked by comparing aerodynamic coefficients obtained on the overtaken body between calculations on different meshes. The side force and the yawing moment coefficients obtained on the three meshes, discussed in the section \ref{numdet2}, are shown in figure \ref{mesh_resol} and compared to the experimental data.\\

\noindent As can be seen, the coarse mesh involves a systematical overestimation of amplitudes, especially on the side force, while the middle and the fine mesh yield similar results: relatively close to the experimental data. For the side force, the critical positions, corresponding to the minimum and the maximum value at $X/L=0$ and $X/L=1$ are in good agreement between the middle and the fine grids. For the first peak of side force, occurring at X/L=1, the maximum difference between the experimental data and the numerical results is +5\% for the fine grid and +7\% for the middle grid. This difference goes up to 21\% for the coarse grid. For the highest peak of yawing moment, occurring at X/L=0.5, the three grids have good agreement with the experimental data with differences of +5\%, +1\% and -5\% for the coarse, the middle and the fine grids, respectively. It is shown that the agreement between the numerical results of the middle grid and the fine grid is very good, proofing the grid convergence.

\subsection{Relative velocity effects}\label{rve}

\begin{figure}[!ht] 
\centering 
\subfigure[$k=0.248$]{
 \includegraphics[width=\textwidth,bb=50 305 725 575,clip=true]{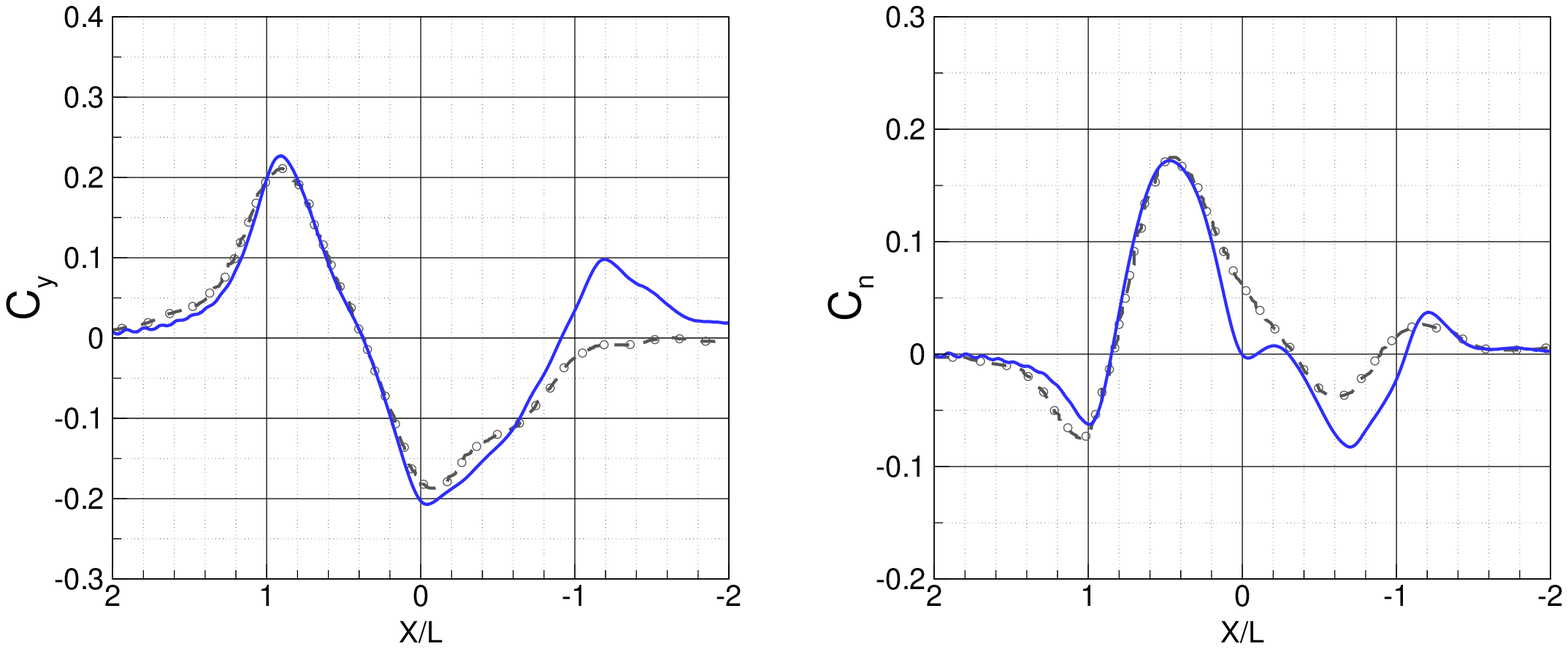}}
\subfigure[$k=0.141$]{
 \includegraphics[width=\textwidth,bb=50 305 725 575,clip=true]{cy_cn_141.eps}\label{141}}
\caption{Relative velocity effects on the overtaken body. Side force (left) and yawing moment (right) coefficients. (\put(0,4){\line(1,0){5}}\put(1,1.5){$\circ$}\put(6,4){\line(1,0){5}}~~~) experimental, (\put(0,4){\color{myblue}\line(1,0){7}}~~~) numerical.}\label{k_slant}
\end{figure}

\noindent Three cases with velocity ratios of $k=0.141$ ($V_{r}=5~m.s^{-1}$), $k=0.248$ and $k=0.331$ ($V_{r}=15~m.s^{-1}$) were simulated. Comparisons of the results obtained for the two relative velocities $k=0.141$ and $k=0.248$ are shown in figure \ref{k_slant} with the experimental data.\\

\noindent For both cases, the numerical results are in good agreement with the experimental data. Nevertheless, the overestimation of the first peak of side force by the numerical simulation is slightly larger for $k=0.141$ than for $k=0.248$. For $k=0.141$, the numerical result of side force is constant between X/L=0 and X/L=-0.5, see figure \ref{141}. However, it is consistent with the experimental data.\\

\begin{figure}[!ht] 
\centering 
 \includegraphics[width=\textwidth,bb=50 30 725 575,clip=true]{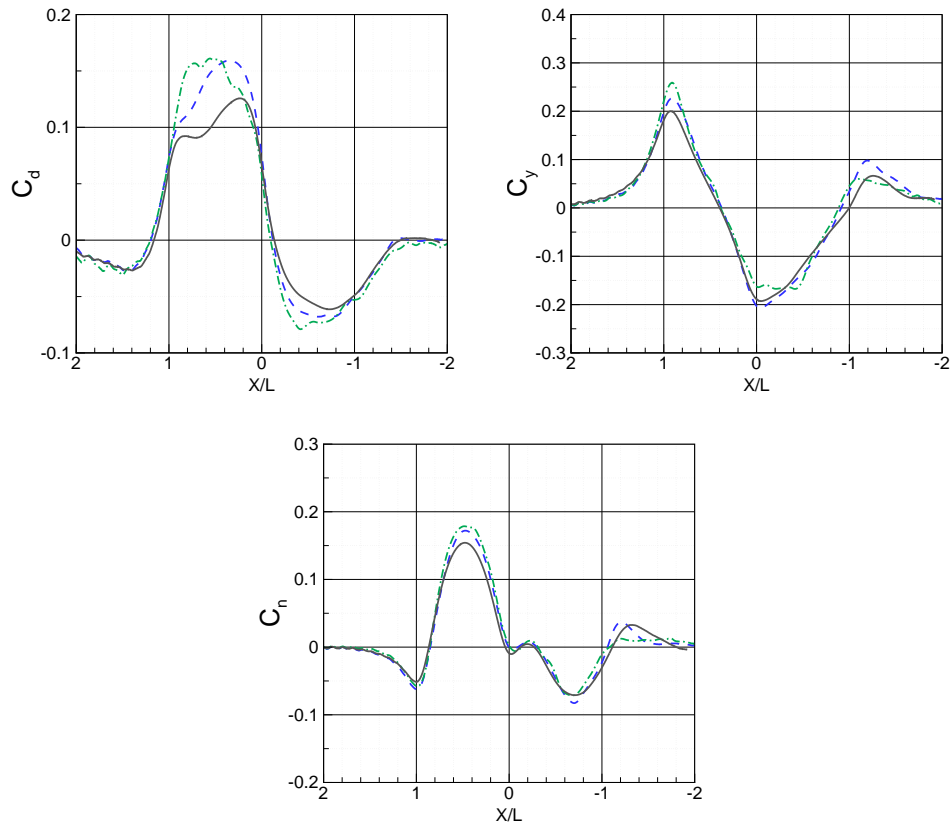}
\caption{Relative velocity effects on the overtaken body. Drag force (top left), Side force (top right) and yawing moment (bottom) coefficients. (\put(0,4){\line(1,0){7}}~~~) $k=0.331$, (\put(0,4){\color{myblue}\line(1,0){5}}\put(6,4){\color{myblue}\line(1,0){5}}~~~) $k=0.248$, (\put(0,4){\color{mygreen}\line(1,0){4}}\put(6,4){\color{mygreen}\line(1,0){1}}\put(8,4){\color{mygreen}\line(1,0){4}}~~~~) $k=0.141$. Numerical results.}\label{3V_comp}
\end{figure}

\noindent Figure \ref{3V_comp} shows comparisons of the numerical results for the three relative velocities. The drag force coefficient presented here, is subtracted from the steady drag coefficient value.\\

\noindent The steady drag coefficient value measured of 0.38 gave a drag increase up to 50\%. This shows that the overtaking manoeuvre can yield a dramatical effect on the car aerodynamic and, therefore, on its fuel consumption.\\

\noindent The existing literature is divided in how the relative velocity of the vehicle influences the aerodynamic coefficients. Noger et al \cite{Noger2004,Noger2005} found that the aerodynamic coefficients are independent of the relative velocity. Corin et al \cite{Corin2008} found that when the relative velocity increased, the drag coefficient increased and the side force decreased. Clarke and Filippone \cite{Clarke2007}, and Gilli\'eron and Noger \cite{Gillieron2004}, shown that an increase in relative velocity yields an increase in the peak coefficients. However, the coefficients are normalized with the velocity of the overtaken vehicle in \cite{Gillieron2004,Clarke2007}. \\

\noindent In the present study, the coefficients decrease when the relative velocity increases.
When the relative velocity increases, the dynamic pressure increases. Therefore, the resulting forces occurring on the overtaken vehicle become more substantial. But these forces are normalized with the overtaking vehicle velocity which takes into account the relative velocity. This can explain the decrease of aerodynamic coefficients peaks.

\FloatBarrier

\subsection{Transverse spacing effects}\label{tve}

In order to study the effects of the transverse spacing, the calculation $k=0.248$ was carried out with two additional transverse spacings: $Y=0.5W$ and $Y=0.7W$. To perform these new calculations, a new mesh was made to take into account the substantial difference of transversal spacing between the case $Y=0.25W$ and the case $Y=0.5W$. This new mesh was deformed, in the transversal direction, for the calculation of the case $Y=0.7W$. \\

\begin{figure}[!ht] 
\centering 
\subfigure[$Y=0.5W$]{
 \includegraphics[width=\textwidth,bb=50 305 725 575,clip=true]{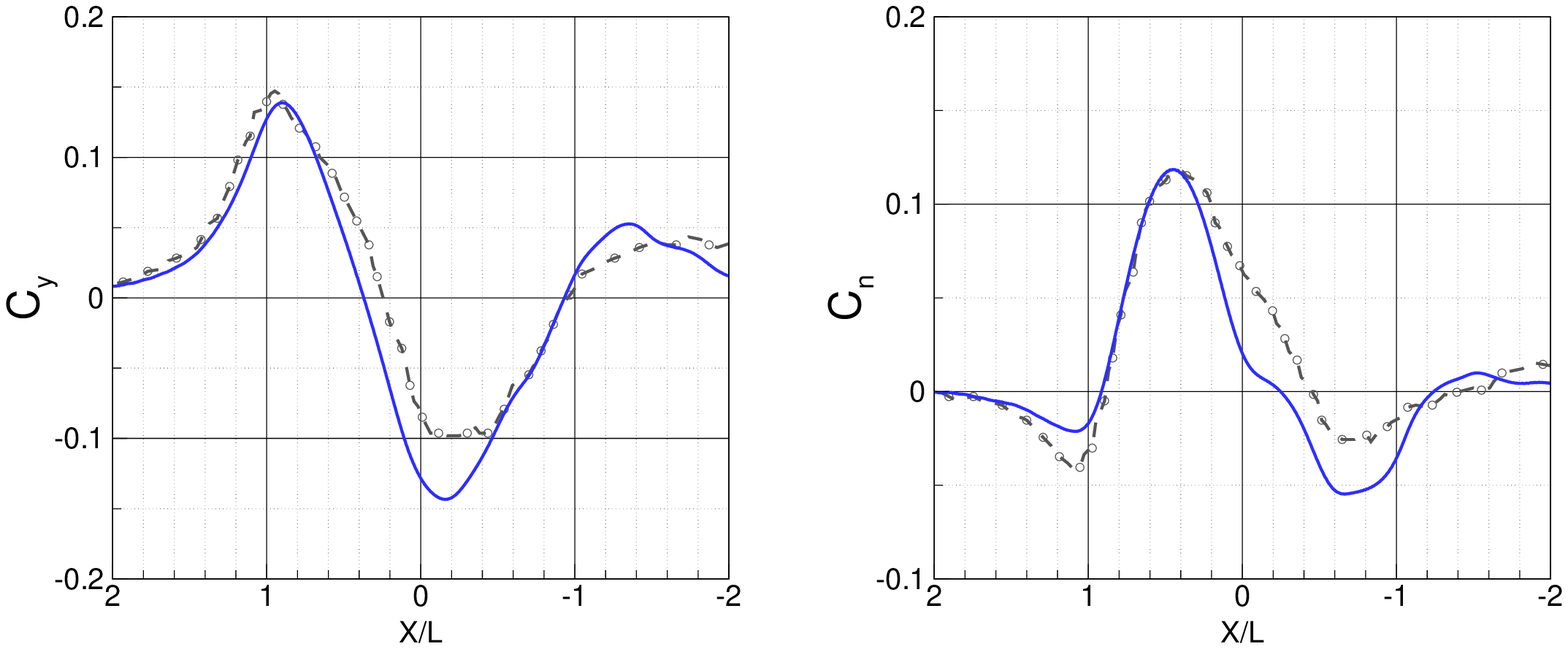}}
\subfigure[$Y=0.7W$]{
 \includegraphics[width=\textwidth,bb=50 305 725 575,clip=true]{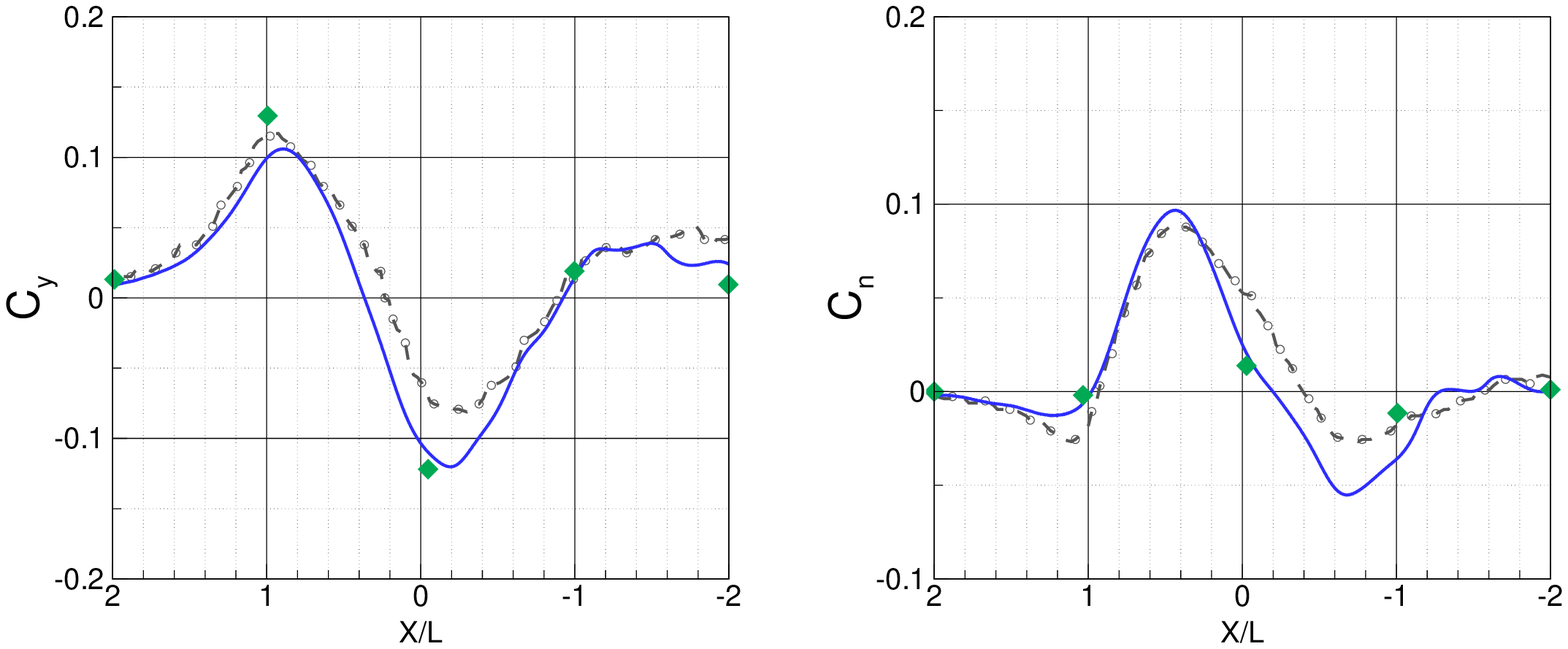}\label{y_spacing_b}}
\caption{Transverse spacing effects on the overtaken body. Side force (left) and yawing moment (right) coefficients. (\put(0,4){\line(1,0){5}}\put(1,1.5){$\circ$}\put(6,4){\line(1,0){5}}~~~) experimental, (\put(0,4){\color{myblue}\line(1,0){7}}~~~) numerical. For $Y=0.7$, (\put(1,0){\color{mygreen}$\Diamondblack$}~~~) quasi-steady numerical results of Gilli\'eron and Noger \cite{Gillieron2004}.}\label{y_spacing}
\end{figure}

\noindent The results, for the side force and the yawing moment coefficients, are shown in figure \ref{y_spacing}. For $Y=0.7$, the quasi-steady numerical results obtained by Gilli\'eron and Noger \cite{Gillieron2004} were available and were added in figure \ref{y_spacing_b}. The experimental data of the quasi-steady case are not shown here. However, these data are very close to the experimental data of the dynamic case. As said previously, the aerodynamic coefficients are independent to the relative velocity in Noger's works, indeed. \\

\noindent Both numerical results are in good agreement with the experimental data. Nevertheless, the numerical simulation underestimates the minimum value of the side force. \\

\begin{figure}[!ht] 
\centering 
 \includegraphics[width=\textwidth,bb=50 30 725 575,clip=true]{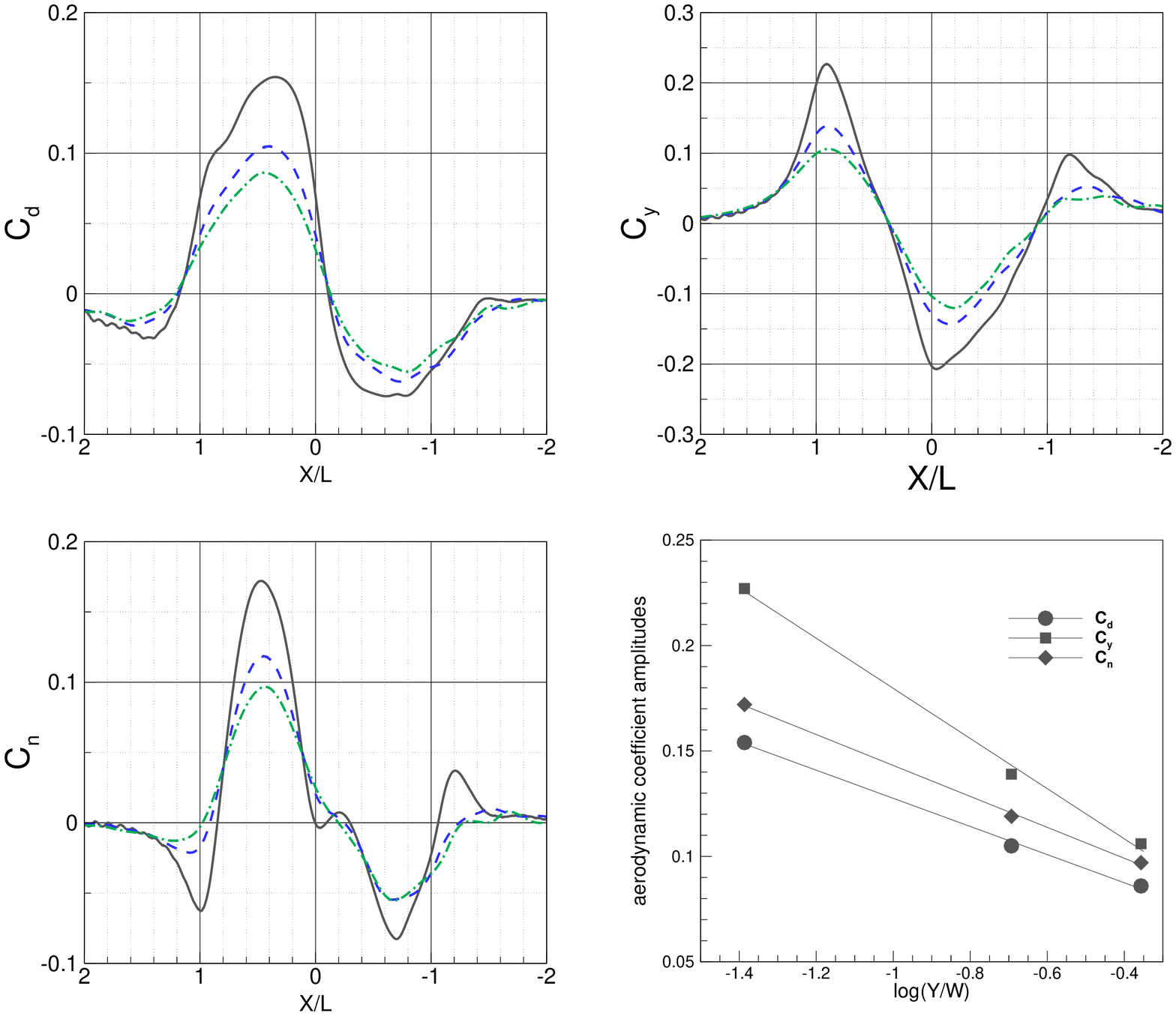}
\caption{Transverse spacing effects on the overtaken body. Drag force coefficient (top left), side force coefficient (top right), yawing moment coefficient (bottom left) and coefficients magnitudes (bottom right). (\put(0,4){\line(1,0){7}}~~~) $Y=0.25W$, (\put(0,4){\color{myblue}\line(1,0){5}}\put(6,4){\color{myblue}\line(1,0){5}}~~~) $Y=0.5W$, (\put(0,4){\color{mygreen}\line(1,0){4}}\put(6,4){\color{mygreen}\line(1,0){1}}\put(8,4){\color{mygreen}\line(1,0){4}}~~~~) $Y=0.7W$.}\label{3Y_comp}
\end{figure}

\noindent Figure \ref{3Y_comp} shows the evolution of the drag force coefficient, the side force coefficient and the yawing moment coefficient for the three different spacings: $Y=0.25W$, $Y=0.5W$ and $Y=0.7W$. The last graph represents the evolution of the coefficient magnitudes as a function of the logarithm of the transverse spacing.\\
It can be easily seen that the coefficient amplitudes reduce when the transverse spacing increases. Besides, the effects occurring at X/L=0 and X/L=-1 are lower for the two highest spacings. The last graph shows that the evolution of magnitudes is a linear function of the logarithm of the transverse spacing.
This result was already shown by Noger et al \cite{Noger2005}.\\

Figure \ref{y_spacing_b} shows that the numerical quasi-steady approach had a tendency to overestimate the magnitude of the side force. The study of the yawing moment coefficient shows that the main phenomena were missed by the quasi-steady approach due to the bad choices of the positions studied. These two arguments proove that a dynamic approach is required for the overtaking process.

\FloatBarrier

\section{Discussion of the passing manoeuvre}
\label{sec:discus}

\subsection{Coefficients behavior}

\begin{figure}[!ht] 
\centering 
\subfigure{
 \includegraphics[width=\textwidth,bb=50 30 725 575,clip=true]{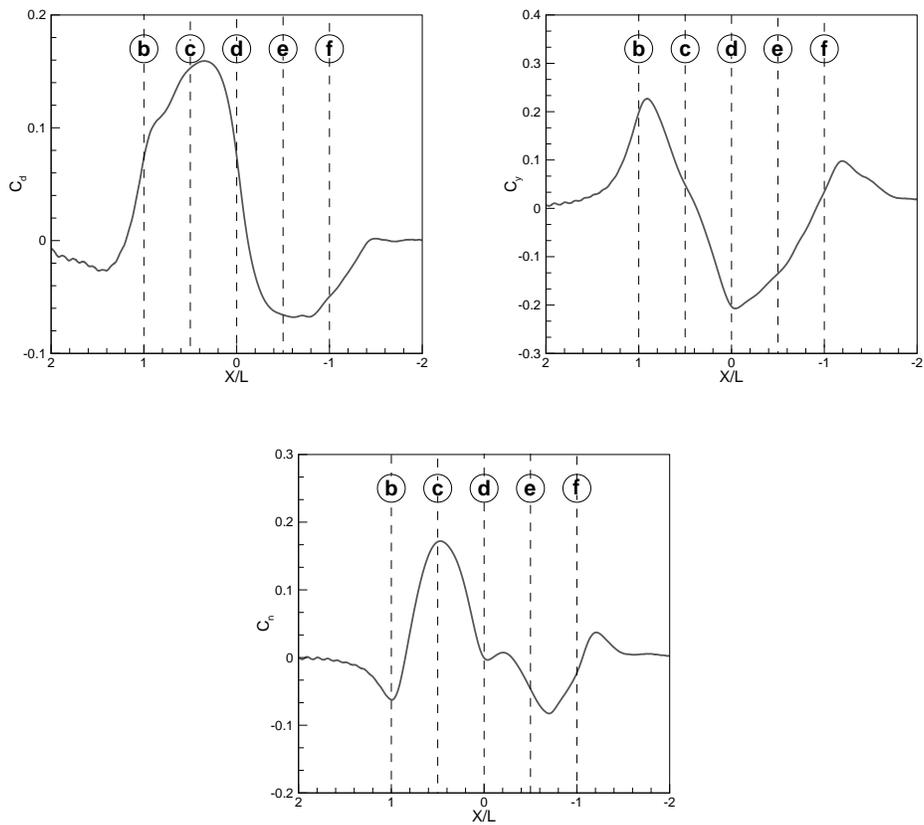}}
\caption{Drag force (top left), side force (top right) and yawing moment (bottom) coefficients.}\label{curve_exp}
\end{figure}

\begin{figure}[!ht] 
\vspace{-3\baselineskip}
\centering 
\subfigure{
 \includegraphics[width=0.6\textwidth,bb=10 125 250 175,clip=true]{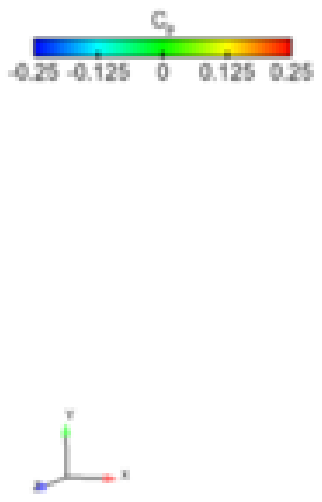}}
\setcounter{subfigure}{0}
\subfigure[$X/L=2$]{
 \includegraphics[width=0.48\textwidth,bb=75 75 275 175,clip=true]{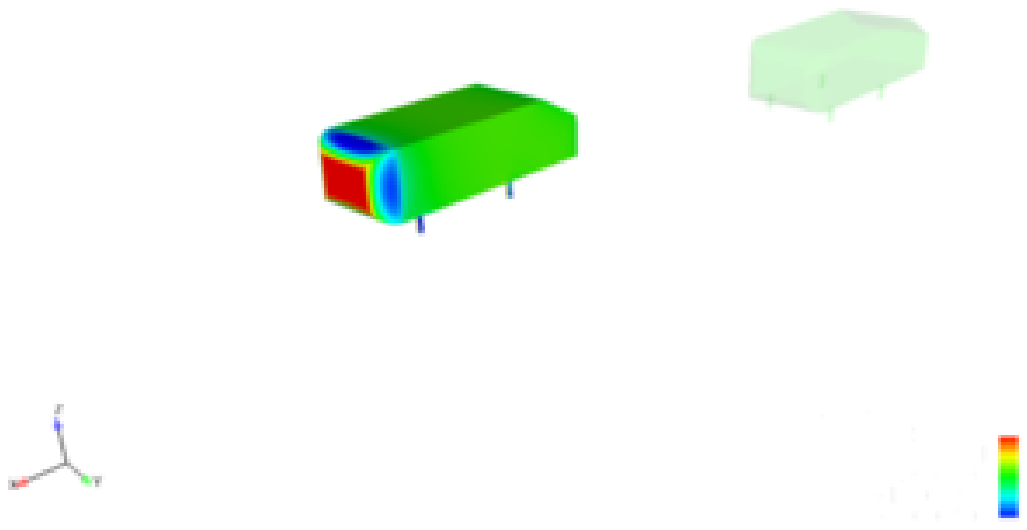}
 \includegraphics[width=0.48\textwidth,bb=0 75 350 225,clip=true]{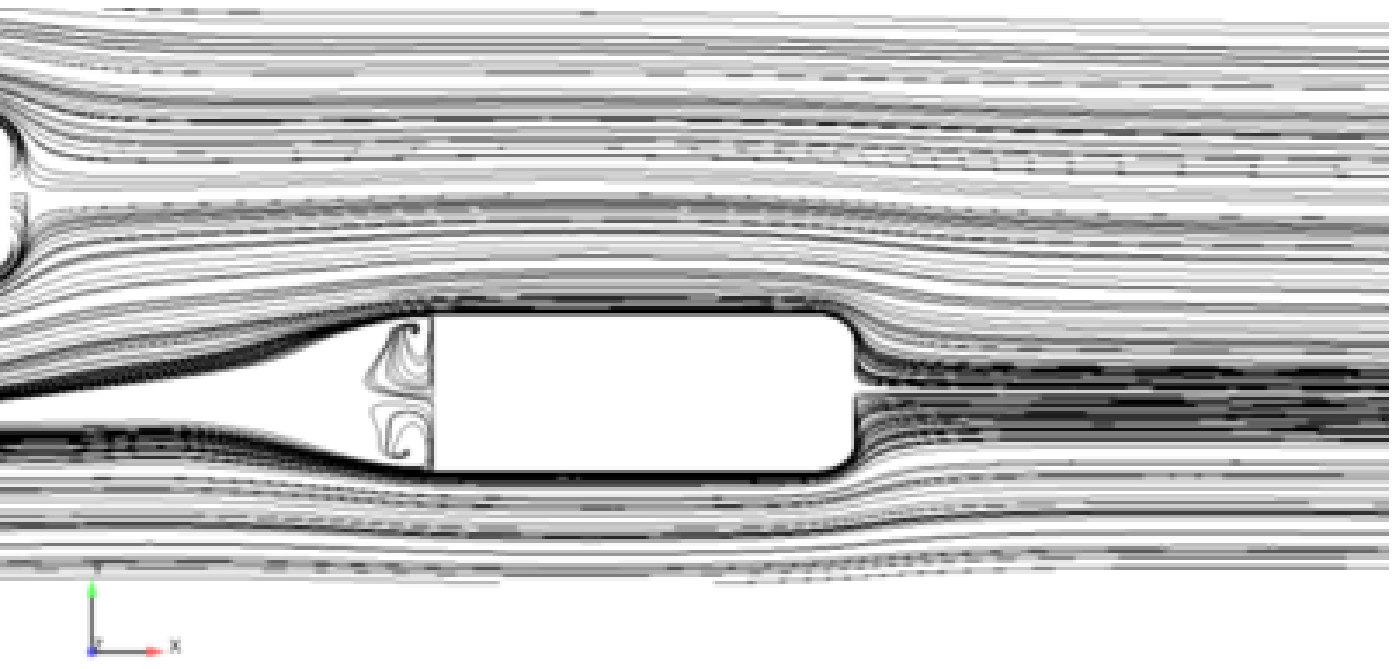}
\label{x=2l}}
\subfigure[$X/L=1$ \textcircled{b}]{
 \includegraphics[width=0.48\textwidth,bb=75 75 275 175,clip=true]{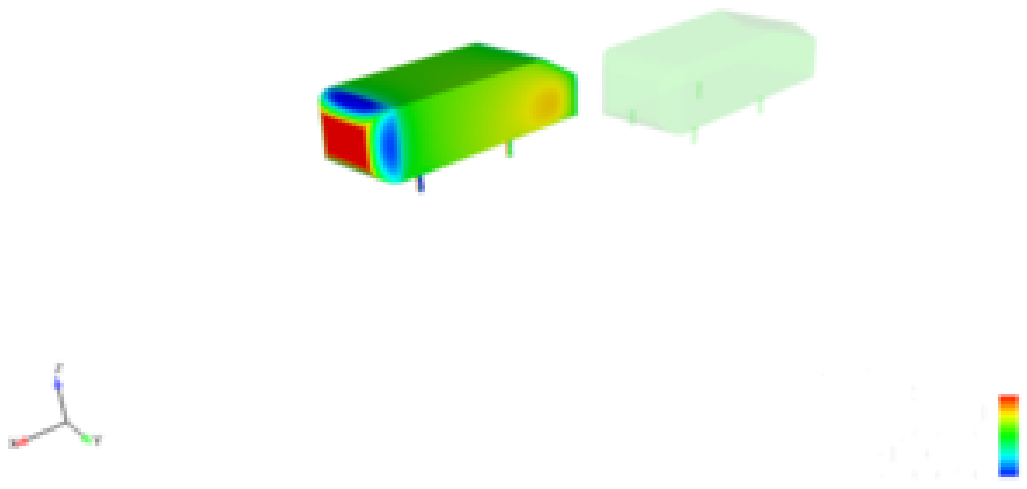}
 \includegraphics[width=0.48\textwidth,bb=0 75 350 225,clip=true]{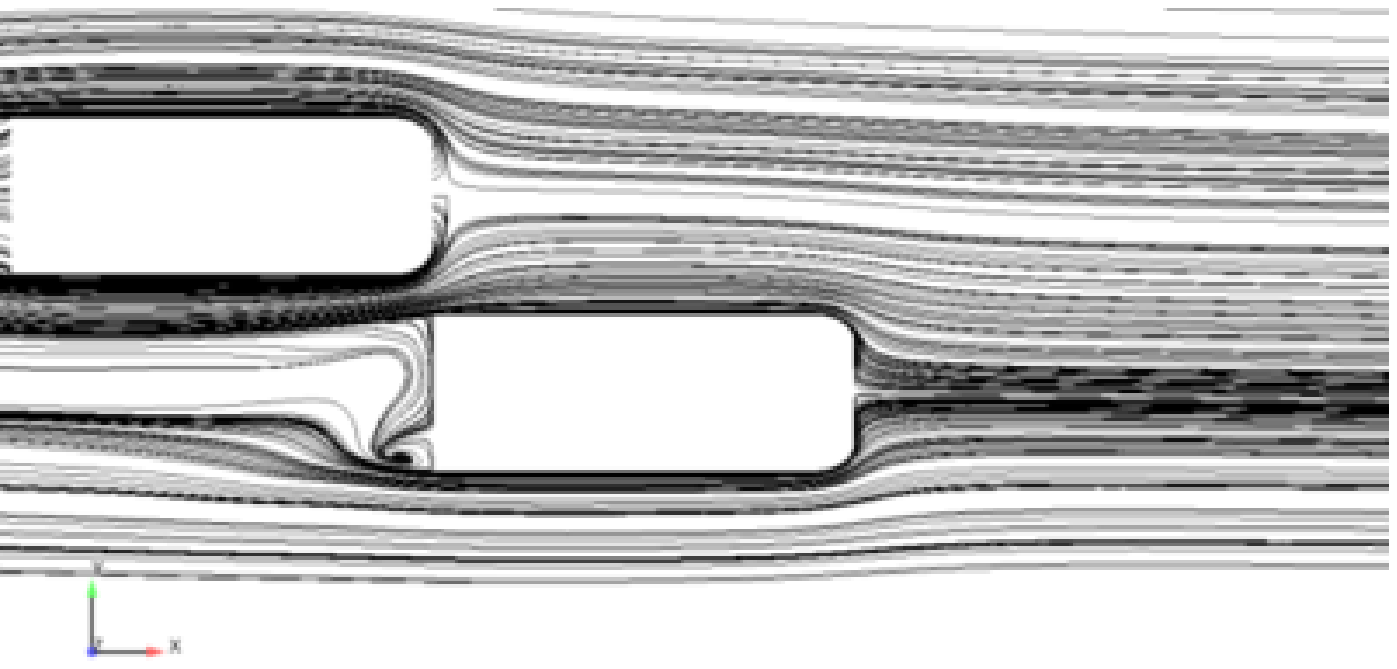}
\label{x=l}}
\subfigure[$X/L=0.5$ \textcircled{c}]{
 \includegraphics[width=0.48\textwidth,bb=75 75 275 175,clip=true]{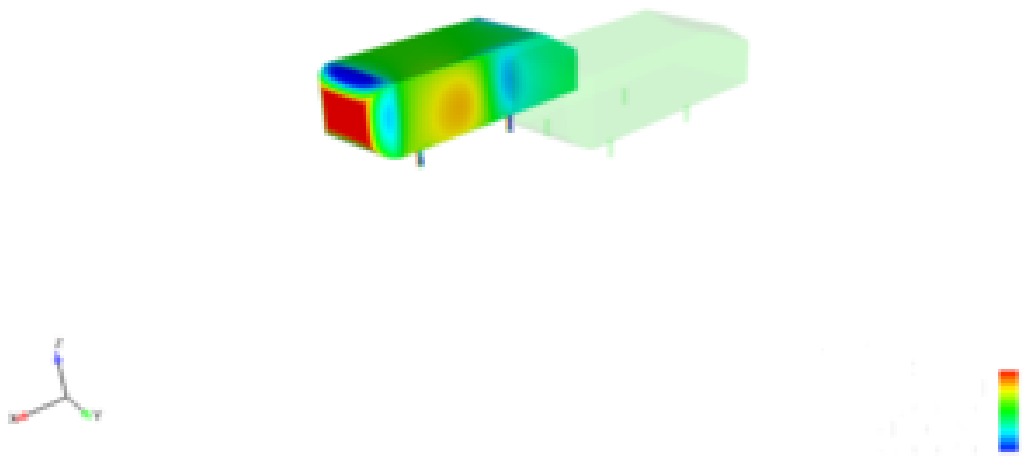}
 \includegraphics[width=0.48\textwidth,bb=0 75 350 225,clip=true]{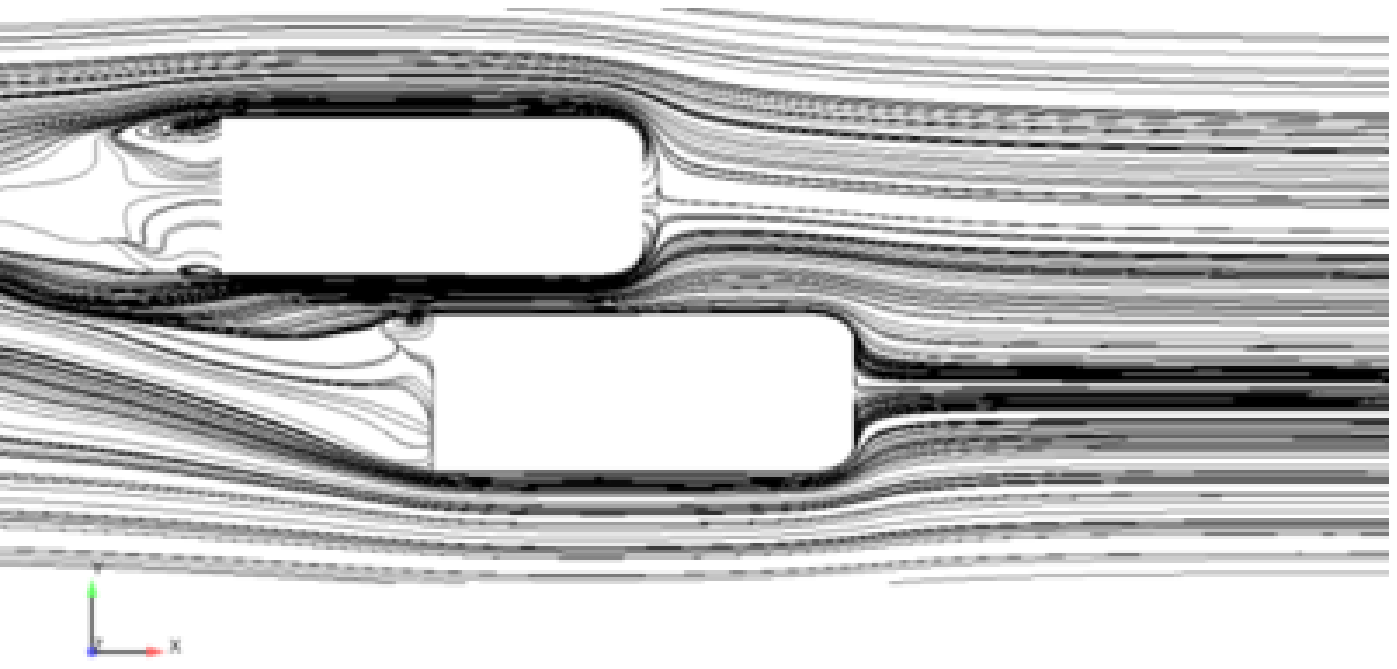}
\label{x=05l}}
\subfigure[$X/L=0$ \textcircled{d}]{
 \includegraphics[width=0.48\textwidth,bb=75 75 275 175,clip=true]{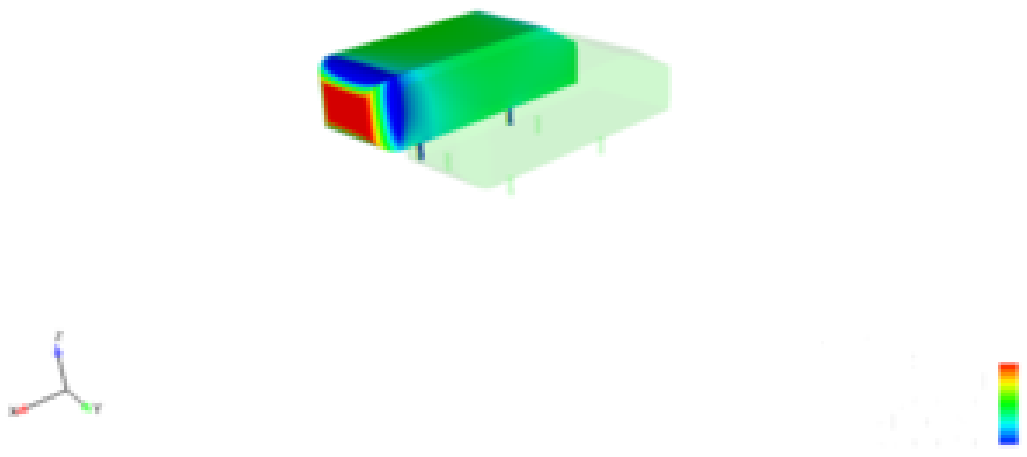}
 \includegraphics[width=0.48\textwidth,bb=0 75 350 225,clip=true]{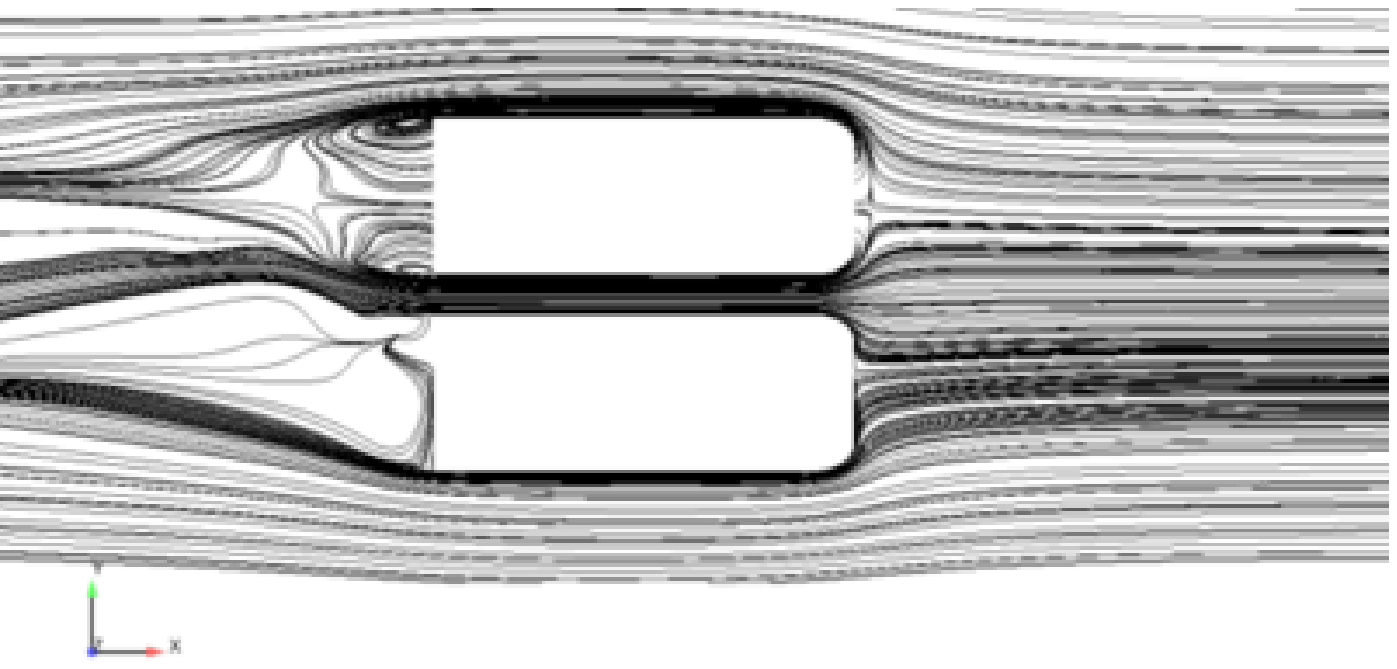}
\label{x=0l}}
\subfigure[$X/L=-0.5$ \textcircled{e}]{
 \includegraphics[width=0.48\textwidth,bb=75 75 275 175,clip=true]{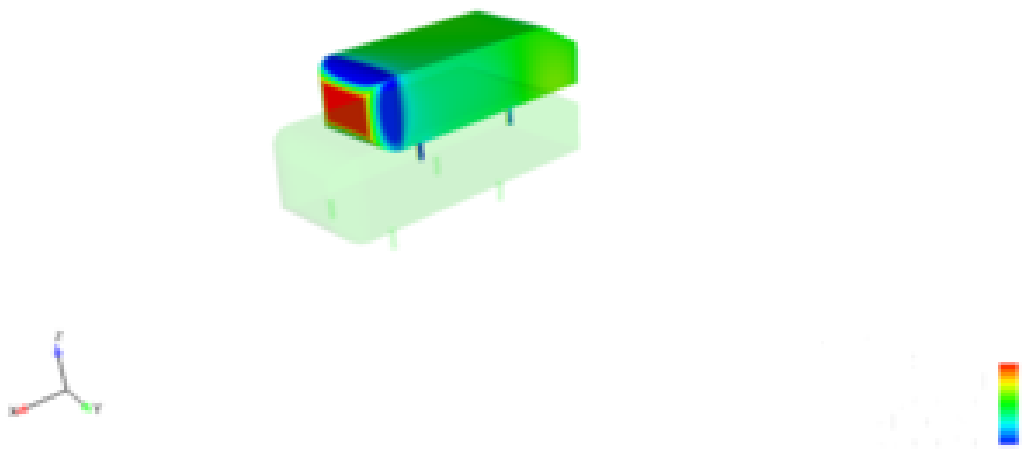}
 \includegraphics[width=0.48\textwidth,bb=0 75 350 225,clip=true]{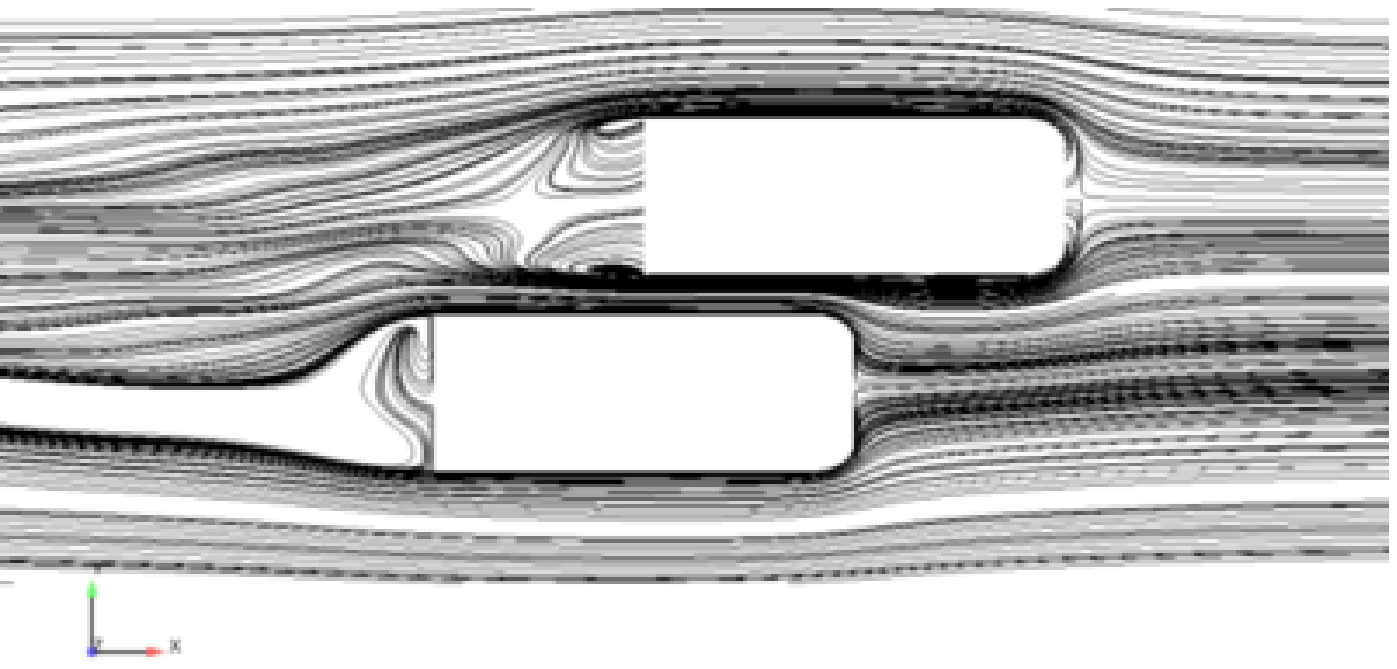}
\label{x=-05l}}
\subfigure[$X/L=-1$ \textcircled{f}]{
 \includegraphics[width=0.48\textwidth,bb=75 75 275 175,clip=true]{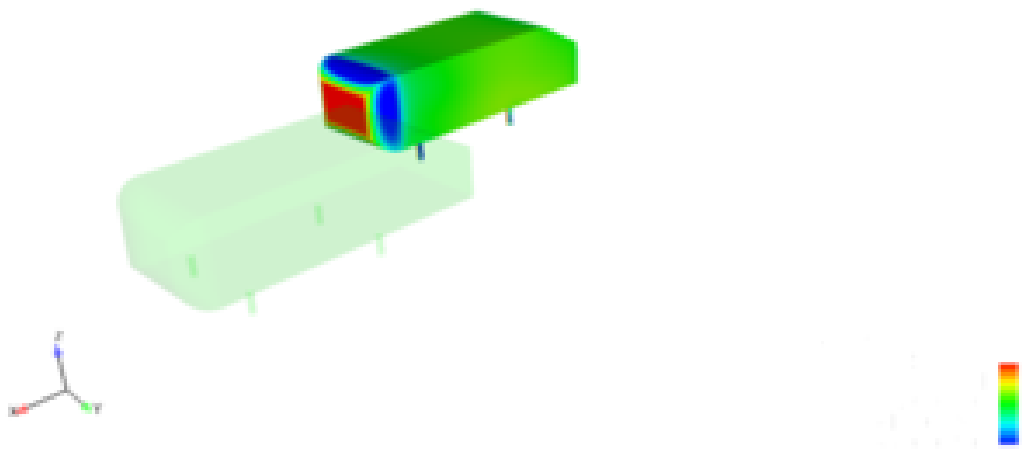}
 \includegraphics[width=0.48\textwidth,bb=0 75 350 225,clip=true]{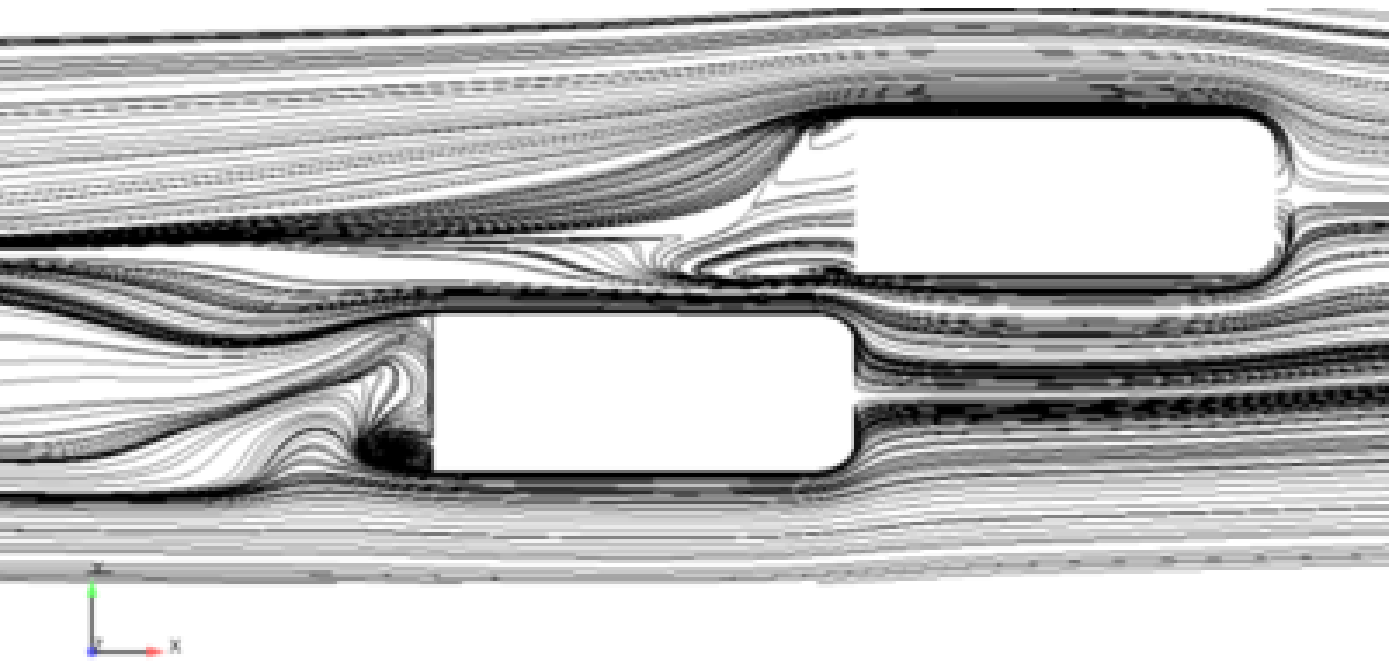}
\label{x=-l}}
\caption{Left: Surface pressure coefficient on the overtaken vehicle. Right: Streamlines of the velocity field projected on the plane $z=0.1558~m$.}
\label{isop_stream}
\end{figure}

Figure \ref{curve_exp} shows the numerical results obtained for the drag force, the side force and the yawing moment coefficients for the reference case $k=0.248$ and $Y=0.25W$. On these three graphs, the 5 vertical dashed lines, labelled by \textcircled{b}, \textcircled{c}, \textcircled{d}, \textcircled{e} and \textcircled{f}, correspond to the critical locations for which the changes are substantial and which require explanations. These labels were chosen to coincide with the ones of figure \ref{isop_stream}.

\FloatBarrier

\noindent The evolution of the pressure coefficient, $C_{p}=2(p-p_{\infty})/\rho V_{\infty}^{2}$ where $p-p_{\infty}$ is the relative pressure, on the inner side of the overtaken vehicle can bring a first explanation. The pressure distributions are shown in the left hand column of figure \ref{isop_stream} for the five moments corresponding to the vertical lines of figure \ref{curve_exp} and for the position $X=2L$ which shows the steady state pressure, i.e. without the influence of the overtaking vehicle. The right hand column of figure \ref{isop_stream} presents the instantaneous streamlines of the velocity field projected onto the plane $z=0.1558~m$, half of the total height of the vehicle, for the six previous positions.

\subsubsection{Flow between $X/L=2$ and $X/L=1$ $\left(\textcircled{b}\right)$} 
The drag force slightly increases before it experiences a sudden pulse. The high pressure in front of the overtaking body increases the pressure at the aft of the overtaken body yielding a drag reduction. After that, the negative pressure in the front side of the overtaking body reduces the pressure at the aft of the overtaken body. The pressure is reduced more and more because the narrowing of the space between both bodies involves an acceleration of the flow. This pressure decrease at the aft of the overtaken body explains the drag pulse at $X/L=1$ $\left(\textcircled{b}\right)$ in figure \ref{curve_exp}.

\noindent The side force increases. As shown in figure \ref{x=l} (left), after its action on the aft of the overtaken body, the high pressure in front of the overtaking body occurs on the inner side of the overtaken body. The result is to increase the side force.

\noindent The yawing moment decreases. The effect of the high pressure on the inner side of the overtaken body is, at first, concentrated on the rear of the overtaken body, as shown in figure \ref{x=l} (left). Then, only the rear of the overtaken body is repelled.  Besides, the approaching of the overtaking body has an effect at the aft of the overtaken body. As can be seen in figure \ref{x=l} (right), the flow separation from the overtaken body's rear end is influenced and the recirculating flow to the aft outer surface is shifted. This change of the flow yields a further anticlockwise moment.

\subsubsection{Flow between $X/L=1$ $\left(\textcircled{b}\right)$ and $X/L=0.5$ $\left(\textcircled{c}\right)$} 

The drag force still increases because a low pressure still acts at the aft of the overtaken body.

\noindent The side force still increases slightly after \textcircled{b} before it decreases. The high pressure further increases the pressure on the inner side of the overtaken body involving the slight increase. When the front of the overtaking body overtakes the rear of the overtaken body, just after \textcircled{b}, the low pressure on the front side of the overtaken body reduces the pressure on the inner side of the overtaken body. This low pressure is added to a Venturi effect, which appears when bodies are close. This global low pressure effect is clearly visible in figure \ref{x=05l} (left). The reduction of the pressure induced explains the decrease of side force.

\noindent The reduction of the pressure on the inner side of the overtaken body is limited to the rear half, as shown in figure  \ref{x=05l} (left), then the rear of the overtaken body is pulled into the path of the overtaking body. Therefore, the yawing moment increases. Moreover, the high pressure is now acting on the front half of the overtaken body, as shown in \ref{x=05l} (left), repelling the front part of this body. As the front half of the overtaken body is repelled, the yawing moment further increases. The combined effects of the low pressure on the rear half of the overtaken body, and the high pressure on the front half of the overtaken body lead to the maximum value for the yawing moment in \textcircled{c}, figure \ref{x=05l} (left).

\subsubsection{Flow between $X/L=0.5$ $\left(\textcircled{c}\right)$ and $X/L=0$ $\left(\textcircled{d}\right)$} 

The drag force slightly increases before it decreases. The low pressure effect at the aft of the overtaken body reduces.

\noindent The side force still decreases and is now negative. The overtaking body passed the central region of the overtaken one, the low pressure effect is now predominating, as shown in figure \ref{x=0l} (left), and the overtaken body is further pulled towards the overtaking body. The effect of the low pressure on the inner side of the overtaken body is maximum when vehicles are side by side, at the position $X/L=0$ $\left(\textcircled{d}\right)$. Therefore, the side force reaches its minimum value at this position.

\noindent The yawing moment decreases. After \textcircled{c}, the low-pressure effect also acts on the front half of the overtaken body, as shown in figure \ref{x=0l} (left), which means that the nose of the overtaken body is pulled into the path of the overtaking one.

\subsubsection{Flow between $X/L=0$ $\left(\textcircled{d}\right)$ and $X/L=-0.5$ $\left(\textcircled{e}\right)$} 

The drag force still decreases and the value of the coefficient is lower than the steady value. The front of the overtaking body overtakes the front of the overtaken body and the low pressure of the front side of the overtaking body decreases the pressure at the fore of the overtaken body. Therefore, the drag decreases.

\noindent The side force increases. The low pressure effect, on the inner side of the overtaken body, reduces. Both vehicles begin to repel each other.

\noindent The yawing moment slightly increases, after \textcircled{d}, before it decreases again. The yawing moment increases slightly because there is an interaction between the flow separations occurring at the aft edges of both bodies, figure  \ref{x=05l} (right). This increase is not found in the experimental data for $k=0.248$, see figure \ref{mesh_resol}. However, it is visible for lower relative velocity $k=0.141$, as shown in figure \ref{141}. After that, the low pressure effect is now acting almost on the front half of the overtaken body, then the front half of this body is pulled into the path of the overtaking body and the yawing moment decreases. 

\subsubsection{Flow between $X/L=-0.5$ $\left(\textcircled{e}\right)$ and $X/L=-1$ $\left(\textcircled{f}\right)$} 

The drag force remains constant to approximately $X/L=0.8$ before it increases. All the effects produced by the overtaking vehicle diminish and the drag coefficient returns to its steady value.

\noindent The side force still increases. Indeed, the low pressure further decreases.

\noindent The yawing moment experiences a slight decrease before it increases again. 

\subsubsection{Flow between $X/L=-1$ $\left(\textcircled{f}\right)$ and $X/L=-2$} 

The drag force increase until its steady value.

\begin{figure}[!ht]
\centering
\includegraphics[width=0.8\textwidth]{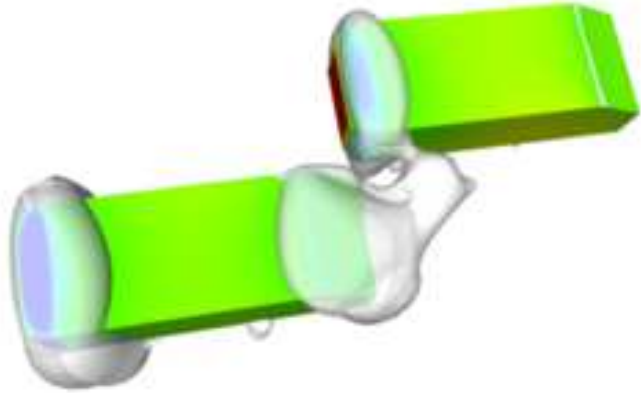}
\caption{Isosurface of pressure, p=20~Pa. At the position X/L=-1.}
\label{isosurface_p}
\end{figure}

\noindent The side force and the yawing moment experience a slight increase before they return to zero. As seen, in figure \ref{isosurface_p}, a positive pressure part, from the rear of the overtaking body propagates towards the overtaken one. The inner sharp edge at the aft of the overtaking body produces a flow separation which repels the overtaken body, as shown in figure \ref{x=-l} (right). Furthermore, the  flow at the aft of the overtaken body, figure  \ref{x=-l} (right), is similar to the flow at the position X/L=1, figure \ref{x=l} (right). The inner flow separation is disrupted by the flow at the aft of the overtaking body and the outer recirculating flow is shifted. This disruption yields to further increase the yawing moment.

\noindent The numerical peak of side force after \textcircled{f}, does not exist in the experimental data, see figure \ref{mesh_resol}. It can be noted that this peak appears on experimental results for lower $k$ in \cite{Noger2004} which are, unfortunately, difficult to consider numerically because of large computational times. This result is inconsistent with the numerical results of Corin et al \cite{Corin2008} for which rounded edges models are used. However, it is consistent with the numerical results of Clarke and Filippone \cite{Clarke2007} for which sharp edges are used. Moreover, Gilli\'eron \cite{Gillieron2003} obtained the same kind of behavior with his simulations of 3D slanted Ahmed bodies. This peak is induced by the flow separation occurring at the inner sharp edge at the aft of the overtaking body. At this point it is not clear if this difference between present prediction and experimental data is related to poor modeling of URANS or to experimental data.

\subsubsection{At $X/L=-2$} 

All the coefficients are returned to their steady values, the overtaking body does not have effect on the overtaken body any more.

\FloatBarrier

\section{Conclusions}
\label{sec:conc}

A three-dimensional numerical methodology, with a deforming/sliding mesh method and the $\zeta-f$ turbulence modelling, was successfully employed to simulate the dynamic passing process between two vehicles. Studies, performed in this work, have highlighted the capacities of the numerical method to well reproduce the effect of the relative velocity and of the lateral spacing on the aerodynamic forces and moments. A complete analysis has enabled to explain all the effects acting on the vehicles.\\
For the overtaken vehicle, it was shown that an increase of transversal spacing involves a decrease of aerodynamic coefficients amplitudes. Moreover, amplitudes of coefficients evolve linearly with the logarithm of the lateral spacing, which confirms the experimental results. Similarly, the aerodynamic coefficients peaks decrease when the relative velocity increases. This conclusion is obviously dependent of the choice of the steady velocity.\\
The future work will extend the present study to overtaking between vehicles with different sizes like a truck and a car. Furthermore, the overtaking in gusty winds will be studied because of safety implications of combination of gusty winds and vehicle overtaking.

\section*{Acknowledgments} This work is supported financially by the Area of Advance Transport at Chalmers. Software licenses were provided by AVL List GMBH. Computations were performed at SNIC (Swedish National Infrastructure for Computing) at the Center for Scientific Computing at Chalmers (C3SE) and National Supercomputer Center (NSC) at Link\"oping University. 

\bibliographystyle{model1-num-names}
\bibliography{postdoc.bib}

\end{document}